\documentclass[11pt,a4paper]{article}
 
\usepackage[margin=1in]{geometry}
\usepackage{amsmath}
\usepackage{amssymb}
\usepackage{graphicx}
\usepackage{bm}
\usepackage{mathtools}
\usepackage{authblk}
\usepackage{cite}
\usepackage{hyperref}
\usepackage{multibib}
\newcites{supp}{References}

\title{Topological Protection of Optical Skyrmions through Complex Media}
\date{}
\author[1,*]{An Aloysius Wang}
\author[1]{Zimo Zhao}
\author[1]{Yifei Ma}
\author[1]{Yuxi Cai}
\author[1]{Runchen Zhang}
\author[1]{Xiaoyi Shang}
\author[1]{Yunqi Zhang}
\author[1]{Ji Qin}
\author[1]{Zhi Kai Pong}
\author[1]{T\'{a}d\'{e} Marozs\'{a}k}
\author[2]{Binguo Chen}
\author[2]{Honghui He}
\author[3]{Lin Luo}
\author[1]{Martin J Booth}
\author[1]{Steve J Elston}
\author[1]{Stephen M Morris}
\author[1,*]{Chao He}
\affil[1]{\small Department of Engineering Science, University of Oxford, Parks Road, Oxford, OX1 3PJ, UK}
\affil[2]{\small Guangdong Research Center of Polarization Imaging and Measurement Engineering Technology, Institute of Biopharmaceutical and Health Engineering, Tsinghua Shenzhen International Graduate School, Tsinghua University, Shenzhen 518055, China}
\affil[3]{\small College of Engineering, Peking University, Beijing 100871, China}
\affil[*]{Corresponding authors: an.wang@stcatz.ox.ac.uk; chao.he@eng.ox.ac.uk}

\renewcommand{\figurename}{Fig.}

\begin{document}
\maketitle

{\bf Optical Skyrmions have many important properties that make them ideal units for high-density data applications, including the ability to carry digital information through a discrete topological number and the independence of spatially varying polarization to other dimensions. More importantly, the topological nature of the optical Skyrmion heuristically suggests a strong degree of robustness to perturbations, which is crucial for reliably carrying information in noisy environments. However, the study of the topological robustness of optical Skyrmions is still in its infancy. Here, we quantify this robustness precisely by proving that the topological nature of the Skyrmion arises from its structure on the boundary and, by duality, is therefore resilient to complex perturbations provided they respect the relevant boundary conditions of the unperturbed Skyrmion. We then present experimental evidence validating this robustness in the context of paraxial Skyrmion beams against different polarization aberrations. Our work provides a framework for handling various perturbations of Skyrmion fields and offers guarantees of robustness in a general sense. This, in turn, has implications for applications of the optical Skyrmion where their topological nature is exploited explicitly, and, in particular, provides an underpinning for the use of Skyrmions in optical communications and photonic computing.} \\

\noindent Since the discovery of the magnetic Skyrmion, there has been a growing interest in the use of topologically non-trivial states in the storage of information \cite{Nagaosa2013, yu_real-space_2010, shi_spin_2021}. In particular, the discretized and particle-like nature of magnetic Skyrmions makes them robust to perturbations \cite{Fert2017, cortes-ortuno_thermal_2017, je_direct_2020} and hence a promising option for future memory \cite{Kang2016} and logic devices \cite{song_skyrmion-based_2020, Zhang2015}. In recent years, Skyrmions have also been predicted and observed in spatially varying polarization fields \cite{Tsesses2018, Gao2020, du_deep-subwavelength_2019, Shen2022, He2022, shen_optical_2023, Bai2020, Lin2021, Shen2021_Super, Cisowski2023, Chao2019, shi_strong_2020, lei_photonic_2021, Liu2022, ye2024theory}, and this has opened the doors to the potential use of such topologically non-trivial states in the transport of information \cite{Shen2023}. In particular, optical Skyrmions have many important properties suited for high-density data applications \cite{Shen2023, shen_optical_2023}, namely (1) the ability to interface with digital information given their integer-valued topological number, (2) the capacity for multiplexing given the independence of spatially varying polarization to other dimensions such as amplitude, phase (including structured phase) and wavelength, and (3) the possibility to store arbitrary large integers within a single localized field. 

Most importantly, there is a compelling heuristic coming from the topological stability of the magnetic Skyrmion that strongly suggests we should expect similar robustness to exist in the optical domain. The argument goes as follows. Given a spatially varying field $\mathcal{S}(x,y)$ taking values in the 2-sphere $S^2$, suppose one can guarantee that the following integral equation \cite{skyrme1961non} evaluates to an integer
\begin{equation}
\label{eq: Skyrmion Number}
    N = \frac{1}{4\pi} \iint \mathcal{S} \cdot \left( \frac{\partial \mathcal{S}}{\partial x} \times \frac{\partial \mathcal{S}}{\partial y} \right) dx dy,
\end{equation}
then, since the integrand in the equation above varies smoothly with respect to $\mathcal{S}$, the mapping $\mathcal{S}\mapsto N$ must also, in somewhat imprecise language, vary smoothly with $\mathcal{S}$. However, as $N$ takes values in the integers, we are left with the remarkable property that smoothly deforming $\mathcal{S}$ cannot change the topological number of the system, for it is impossible to smoothly transition from one integer to another without also passing through every number in between. This then presents an important question on the topological robustness of optical Skyrmions: under what conditions can we guarantee that the above integral equation evaluates to an integer for a given spatially varying polarization field? 

Putting aside integrality for the moment, the topological robustness established by the argument above is the primary motivation for investigating the use of optical Skyrmions as a method of reliably transmitting data through noisy channels, such as in communications and computing. The key property of electromagnetic fields at play here is the ellipticity of the underlying Helmholtz equation, which guarantees the smoothness of the electric field in propagation. This smooth deformation of the electric field along its direction of propagation then descends onto a smooth deformation of polarization fields, which excluding technical subtleties we mention later, establishes a constant topological number in propagation even as the transverse polarization profile changes. Moreover, this conclusion remains largely unchanged for optical fields propagating in media with continuously varying material parameters, which suggests a degree of robustness against aberrations. 

There are, however, several challenges in turning the above into a formal argument, the most problematic being that the Skyrmion number integral is not always guaranteed to be integer-valued in the optical setting \cite{Liu2022, Gao2020, ye2024theory}. To elaborate, magnetic Skyrmions are naturally topological due to a preferred alignment of dipoles in the far field, which we later show to be precisely the correct mathematical condition for the Skyrmion number integral to evaluate to an integer. On the other hand, electromagnetic fields do not have a naturally preferred polarization state at infinity unless specifically designed in that way. Moreover, polarization singularities arising from zeros of the electromagnetic field are also volatile and difficult to control. These factors contribute to the existence of ``fractional'' phenomena in optical fields that otherwise cannot be observed in magnetic spin textures, where the above integral equation evaluates to a non-integer. Such fractional phenomena have important implications for the robustness of the optical Skyrmion, wherein the usual argument for topological stability breaks down \cite{Liu2022}. To date, investigations into the topological robustness of spatially varying polarization fields with integer (and non-integer) Skyrmion numbers are still in their infancy, and there is scope to develop a more sophisticated theory that elucidates the conditions that guarantee topological protection. 

Motivated by the above, we give a precise and purely topological definition of the Skyrmion that clarifies the properties of an $S^2$-valued field that allow for an integer topological number to be defined, and from this, show that the topological character of the Skyrmion is derived from its boundary. We then present a completely general framework for handling perturbations that applies to all $S^2$-valued fields, including optical, magnetic Skyrmions, etc., which builds on recent pioneering topological robustness results established in non-local quantum Skyrmions \cite{ornelas_non-local_2024, ornelas2024topologicalrejectionnoisenonlocal}, and the propagation of optical Skyrmions through homogeneous retarders \cite{he2023universal, teng_physical_2023}. We then specialize our theory to paraxial Skyrmions from which a topological duality between light and matter emerges, namely that the boundary conditions satisfied by the Skyrmion are also the required boundary conditions of material parameters that guarantee topological protection. Lastly, we present experimental results that validate the robustness of the Skyrmion number as the field propagates through a complex optical system containing both typical polarization aberrations from common optical elements and perturbations mimicking more extreme situations in the real world. Our work gives new insight into the topological robustness of the optical Skyrmion and paves the way for their use in high-density data applications such as optical communications and photonic computing.  

\section{Main}

\subsection{Role of the Boundary in Skyrmions}

As mentioned in the introduction, the lynchpin to establishing topological robustness is exhibiting conditions that guarantee the integer-valued nature of the Skyrmion number integral. We demonstrate in Methods 1 and 2 that the boundary conditions of the field determine when this is true, and we refer to fields that can be given topological character in this way {\it compactifiable}. 

\begin{figure}[!ht]
    \centering
    \includegraphics[width=0.96\textwidth]{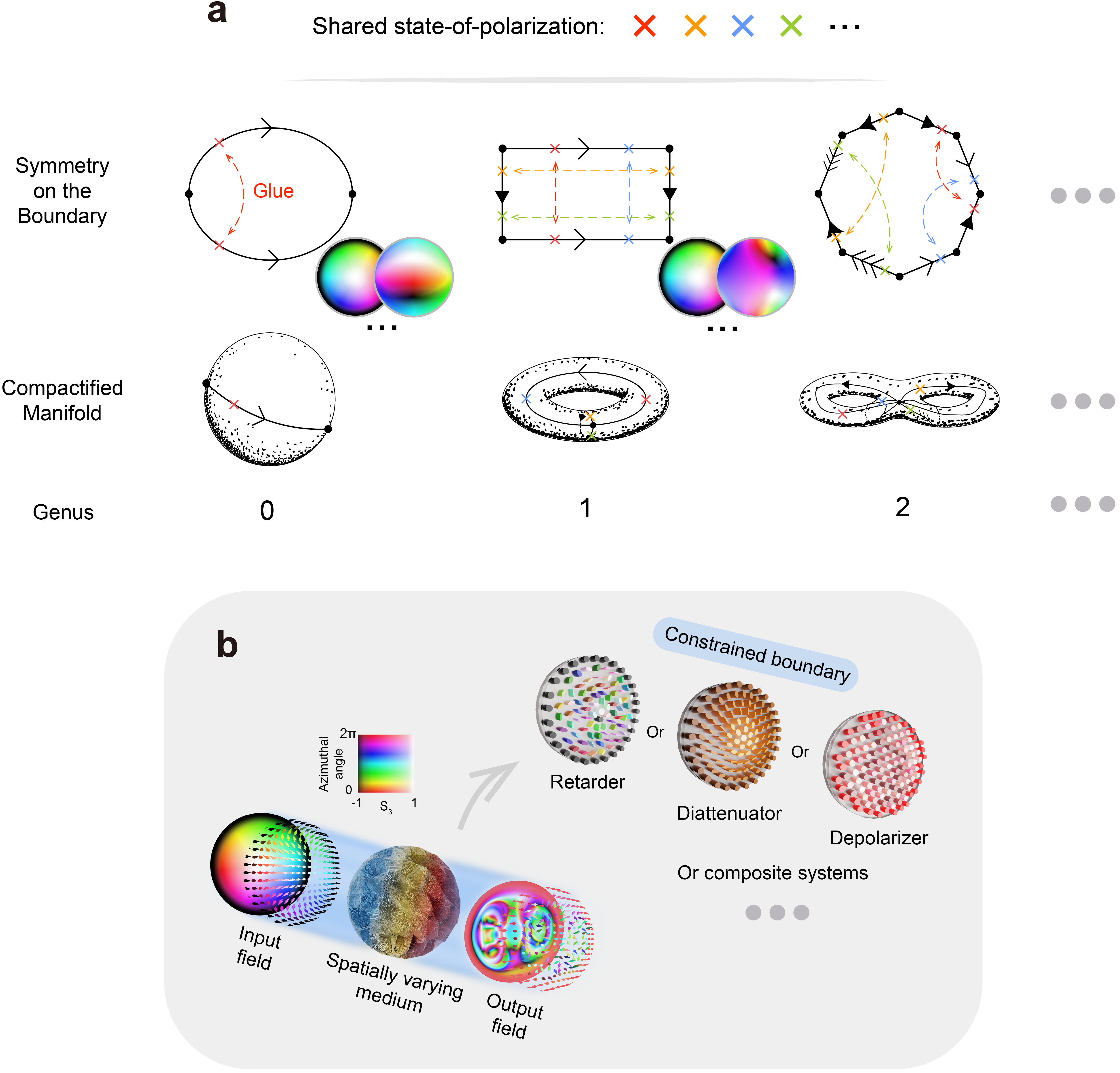}
    \caption{{\bf Definition of Skyrmions and their interaction with various polarisation aberrations.} {\bf a}, The various symmetry properties of the field on its boundary that allow for compactification and the corresponding compactified manifold. The symmetry properties are shown here via polygons, with arrows indicating pairwise identification of different edges oriented in the direction of the arrow. Notice that certain symmetries allow the field to be tesselated, which gives topological character to periodic structured fields. As mentioned in the main text, due to the topological nature of compactification, one can freely deform this polygon and, in the case of unbounded fields, one can consider the boundary at infinity. Possible Stokes fields satisfying the relevant symmetries are also shown. It is worth noting that a field with a constant value on its boundary satisfies every symmetry property simultaneously, however, there exist fields, non-uniform on their boundary, that can also be compactified, as shown in the figure. Throughout this paper, Stokes fields are depicted using hue to signify azimuthal angle $\tan\theta = s_2/s_1$ and saturation to represent height $s_3$ (similar to \cite{shen_optical_2023}). {\bf b}, Schematic depicting the effects of a complex spatially varying medium on an incident optical Skyrmion field. In this paper, systems exhibiting spatially varying retardance, diattenuation, depolarization, and combinations of the aforementioned are carefully considered. Note that the boundary of the incident field and the corresponding aberrations are of key importance in guaranteeing topological protection. Throughout this work, fields with a constant value on their boundaries are considered for physical reasons expanded on in the main text.}
    \label{fig:concept}
\end{figure}

In simple terms, a field is compactifiable if its domain can be viewed as a compact and orientable surface, which via the classification theorem for compact surfaces, is characterized entirely by its genus, or more intuitively, the number of ``holes'' of the surface \cite{kinsey_topology_2012}. From standard polygonal constructions of topological surfaces, one way to understand compactifiability is as a symmetry property of the boundary, where the appropriate symmetries allow one to ``glue'' the boundary together into a surface. Fig. \ref{fig:concept}a gives examples of different symmetries that allow for compactification. 

Notice that while a field constant on its boundary satisfies all possible symmetries, there also exist fields, non-uniform on their boundary, for which compactification holds (See Fig. \ref{fig:concept}a). Moreover, while polygons are used to depict these symmetries, the topological nature of compactification allows deformations of these polygons. In the case of optical fields, which need not necessarily be confined, we can extend our considerations to symmetries of the boundary at infinity.

Notably, if a field $\mathcal{S}\colon U \longrightarrow S^2$ is compactifiable, then it necessarily satisfies an uncountable number of non-trivial integral equations 
\begin{equation}
    \int_U \mathcal{S}^\ast \omega = \deg \mathcal{S}\int_{S^2} \omega,
\end{equation}
one for every $\omega \in \Lambda^2(S^2)$. The integer-valued nature of the usual Skyrmion number integral equation is then a consequence of the above. We also show in Methods 3 that compactifiable fields satisfy a natural notion of homotopy invariance from which the Skyrmion derives its topological robustness. Lastly, note that while the usual integral equation evaluates to an integer for a Skyrmion, the converse need not be true, and just knowing that equation (\ref{eq: Skyrmion Number}) evaluates to an integer for a specified field gives no clear topological information about the underlying field and no clear indication of whether the integral equation is invariant under homotopies of the field. 

With the notion of compactifiability, we can revisit the arguments for the topological robustness of optical Skyrmions presented in the introduction. As before, the ellipticity of the Helmholtz equation provides the necessary regularity for the natural propagation of electric fields to descend onto homotopies of polarization fields. If, further, the polarization field has the necessary boundary conditions to be compactifiable on every transverse plane, then the integer-valued nature of the Skyrmion number integral along these planes combined with the usual smoothness argument once again establishes topological protection in propagation. Notice that our theory is a generalization of the usual methods applied to paraxial Skyrmion beams that rely on a superposition of Laguerre-Gaussian (LG) modes and proves topological protection for any complex superposition of modes provided that there are no zeros of the field and that compactification holds (for example, if the polarization state approaches a constant value at infinity). Moreover, this approach relies on neither a specific modal decomposition, whose linear nature is intrinsically incompatible with the decidedly non-linear nature of topology, nor theoretical computations of Skyrmion numbers along the transverse plane, which quickly become intractable as the number of superposed modes increases. 

We give a more formal description of this procedure for establishing topological robustness in Methods 3. Note also that our approach does not reference any particular property of optical fields and can, therefore, be used to establish the preservation of the Skyrmion number in any number of settings beyond paraxial optical Skyrmions, including optical Skyrmions in evanescent fields, magnetic Skyrmions, etc. Moreover, the procedure we have introduced is, in some sense, both necessary and sufficient in proving topological protection. This is because if we compactify, as we usually do, onto the 2-sphere, then the Hopf degree theorem provides the converse statement that if topological protection holds, then there exists some way of demonstrating this fact using the procedure in Methods 3.  

\clearpage
\subsection{Specializing to Optical Aberrations and the Paraxial Optical Skyrmion}

To further demonstrate the utility of our approach, we specialize our methodology to paraxial optical Skyrmions and show that in this case, the Skyrmion number remains conserved across a broad range of important optical aberrations including spatially varying retarders, diattenuators, specific classes of depolarizing systems, and more generally, any complex system composed of such elements (see Fig. \ref{fig:concept}b for a pictorial representation of the systems under consideration and Methods 4-7 for detailed derivations). Note that we exclude strongly depolarizing and absorbing media in our considerations, which can disrupt topological character by introducing polarization singularities into the field. However, since such aberrations are not prevalent in optical systems, our theory remains applicable to the vast majority of cases, with extensions to more severe aberrations an exciting avenue to explore.  

In each case, we show that the compactifiability of material parameters is one of the conditions needed for ensuring topological protection, with a general tolerance to perturbations away from the boundary provided they remain smooth. Our result is a consequence of the compactification formalism introduced, which makes explicit the fact that the topological nature of the Skyrmion is derived from its boundary. This is particularly important for paraxial optical Skyrmions, whose natural boundary is at infinity due to the infinite extent of the fundamental solution to the Helmholtz equation and, therefore, remains unaffected by the necessarily finite aberrations along the direction of propagation. 

We would also like to emphasize that retarders, diattenuators, and depolarizers are abstract mathematical concepts \cite{chipman_polarized_2018} that can be used to describe polarization aberrations and are formalisms that can be applied to a large number of optical components \cite{chipman_polarized_2018, mcguire_polarization_1994}. Since such polarization aberrations alter the field of a paraxial Skyrmion beam, potentially affecting the topological information carried, it is important to study and quantify the topological robustness of the Skyrmion against such aberrations. 

More concretely, important aberrations that can be described in this way include the Fresnel effect from reflections and refractions, anisotropic absorptions of coatings, birefringence, depolarization induced by imperfections of optical components, etc. The typical levels of such aberrations have been studied before \cite{he_vectorial_2023} and, in the majority of cases, are weak spatially varying smooth aberrations with similar properties on their boundary. This observation links the abstract mathematical notion of compactifiability to physical aberrations in real optical systems and provides a heuristic justification for the choice of aberrating elements and generated Skyrmion beams in the experiments that we describe below. In these experiments, we validate the robustness of the optical Skyrmion against various optical aberrations, including both aberrations from common optical elements as well as atypical aberrations that mimic more extreme perturbations that exist in real-world settings. Lastly, we reiterate that the aberrations considered in this section have been selected due to their physical significance, and are merely specific examples of perturbations that can be handled by the generality of our theory.

\subsection{Experiments}

\begin{figure}[!p]
    \centering  \includegraphics[width=0.95\textwidth]{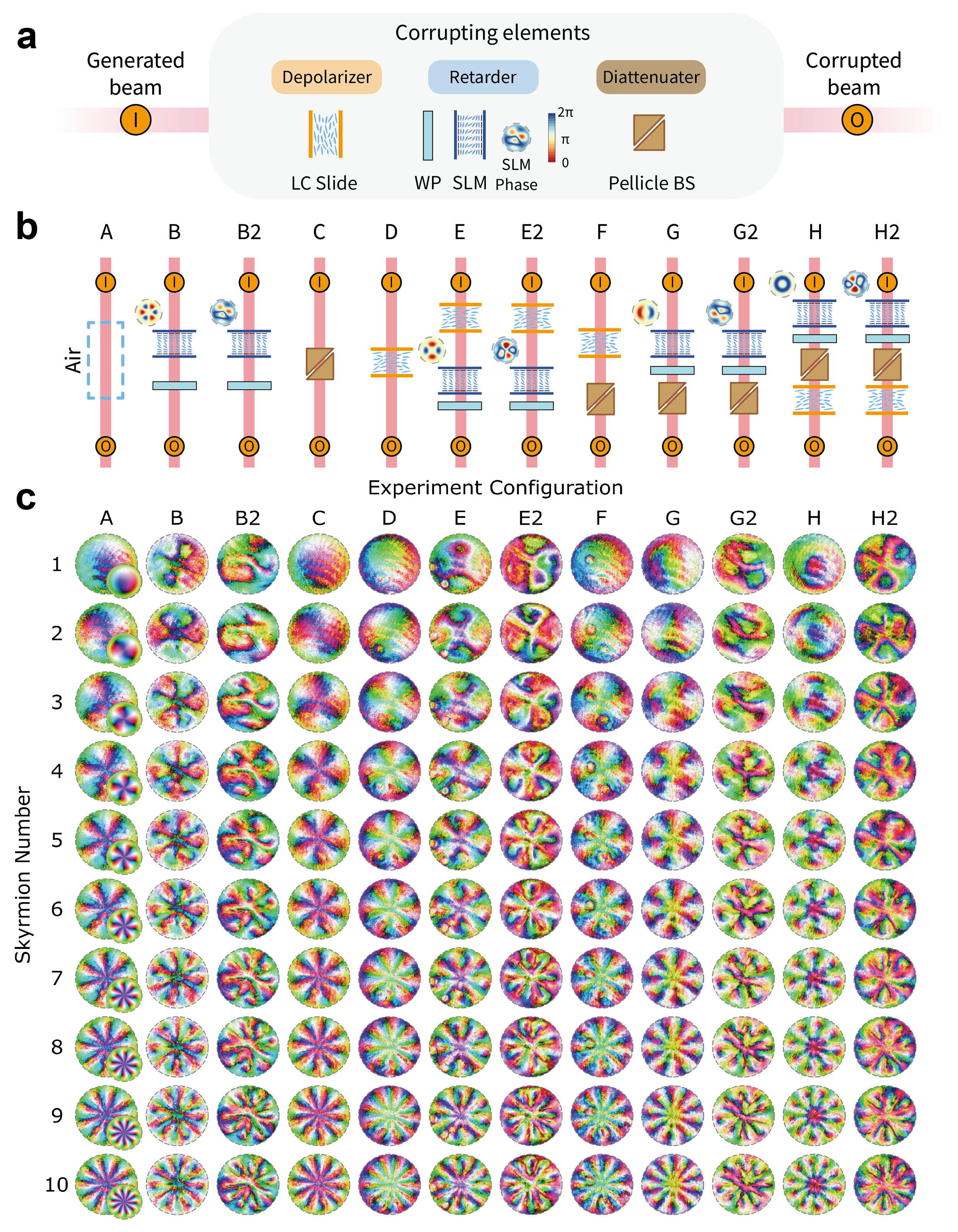}
    \caption{{\bf Experiment assembly and measured Stokes fields.} {\bf a}, The various corrupting elements used in testing, and the class of media they belong to. Here, I indicates the input beam and O the output beam. {\bf b}, The order of components and SLM phase patterns used in testing. {\bf c}, The measured Stokes fields for different experiment configurations and input Skyrmion numbers. Simulated Stokes fields for configuration A are also shown.}
    \label{fig:setup}
\end{figure}

To back up our theoretical work, we present experimental results that validate the robustness of the Skyrmion number as the field propagates through a complex optical system. In our experiments, we synthesized Skyrmions with degrees ranging from 1 to 10 and propagated these Skyrmions through a complex optical system with multiple aberrating elements. We then measured the Stokes field using standard Stokes polarimetry \cite{he_polarisation_2021} and computed the Skyrmion number integral from our measurements using a technique described below. The target Stokes fields generated in our experiments take the form
\begin{equation}
\label{eq: target stokes field}
    \begin{pmatrix}
        S_1 \\ S_2 \\ S_3
    \end{pmatrix} = \begin{pmatrix}
        \sqrt{1-f(r)}\cos n \theta \\ f(r) \\ \sqrt{1-f(r)}\sin n \theta
    \end{pmatrix}, \quad f(r) = \left\{\begin{array}{r l}
         \exp\left(1-\frac{1}{1-r^2} \right)-1, & 0\leq r \leq 1 \\
         -1, & 1<r
    \end{array} \right.,
\end{equation}
where $n$ is the target degree of the generated Skyrmion and $r$ is a normalized radius. This field is realized using a cascade of two liquid crystal spatial light modulators (LC-SLM) aligned at a 45\textsuperscript{$\circ$} angle to each other as shown in Supplementary Fig. 3, and where the same number of pixels is used in generation for all experiments. For a detailed description of the pixelated arbitrary state-of-polarization generation technique adopted, see \cite{hu_arbitrary_2020, he2023universal}. Note also that this particular choice of Stokes field was made as it is everywhere differentiable, even as we transition from the spatially varying part when $r \leq 1$ to the homogeneous one when $r > 1$. As a consequence, we may interpret the generated field as a Skyrmion, in the sense of compactifiability, for any open set $U$ containing the unit ball centered at the origin. Note this boundary condition also matches the material properties of the various aberrations used, which as explained in Methods 4-7, is the duality condition in guaranteeing topological protection. 

\begin{figure}[!ht]
    \centering  \includegraphics[width=0.95\textwidth]{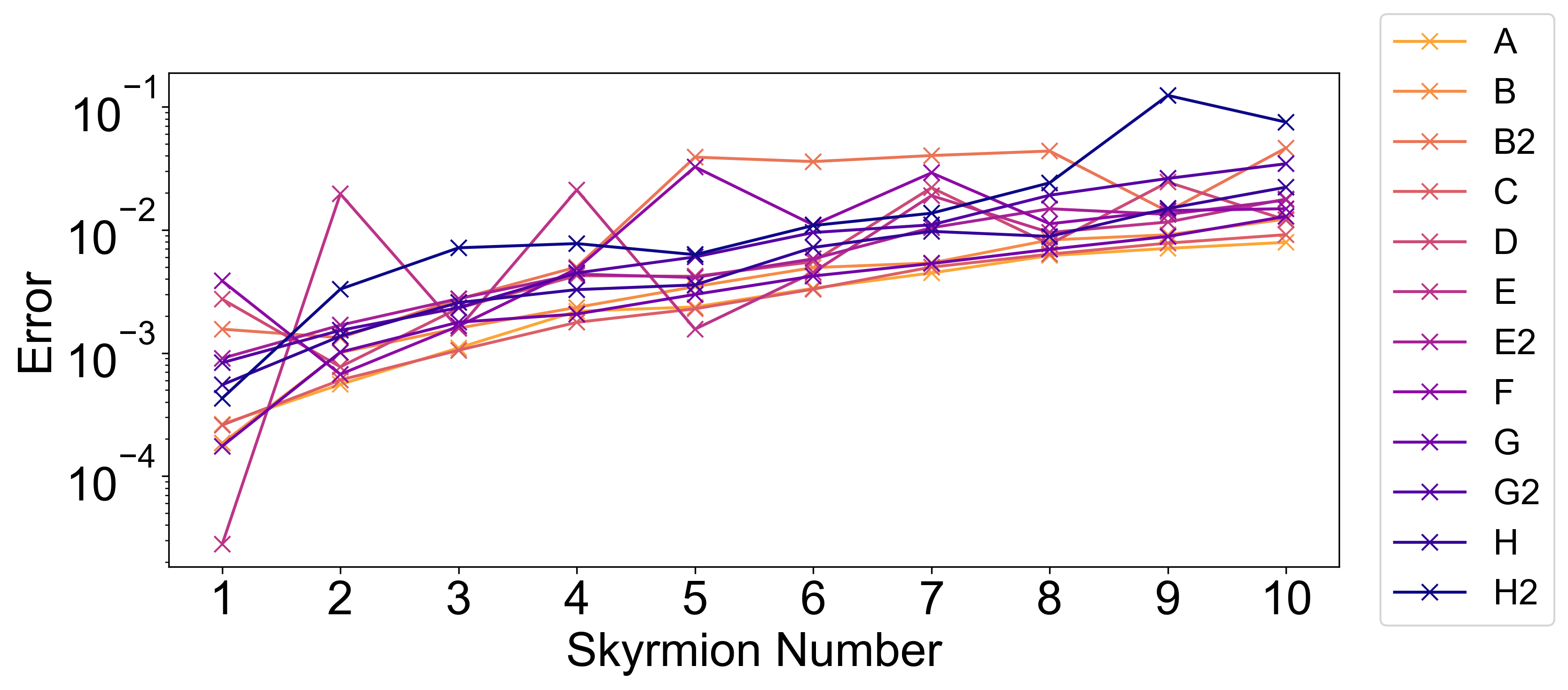}
    \caption{{\bf Computed Skyrmion number and associated error.} The associated error as defined in equation (\ref{eq:experiment_error}) across all configurations.}
    \label{fig:results}
\end{figure}

The corrupting elements in our experiments consist of a cascade of up to four components, including an LC-SLM and variable waveplate that together act as a spatially varying retarder, a pellicle beamsplitter that possesses a high diattenuation of approximately 0.25, and a nematic LC mixture sandwiched between two glass substrates that simultaneously acts as a retarder while also weakly depolarizing incident light (Fig. \ref{fig:setup}a). Importantly, these elements mimic the effects of common aberrations found in optical systems as well as stronger, more atypical aberrations that may exist in the real world. The different aberrating elements and their relative placements in testing (from A to H2) are shown in Fig. \ref{fig:setup}b, the observed Stokes fields measured in experiments shown in Fig. \ref{fig:setup}c, and the error in computed Skyrmion numbers shown in Fig. \ref{fig:results}. We also provide a quantitative assessment of the level of aberrations present in select configurations via Mueller matrices in Supplementary Fig. 4.

We would like to highlight three important technicalities. Firstly, in the synthesis of our fields, a relatively small fraction of the SLM is dedicated to the spatially varying portion of the field while the majority of pixels are used to impose boundary conditions. Due to the homotopy invariance of the degree, the demands on the resolution required to synthesize the spatially varying part of the field are fairly gentle. Indeed, the homotopy class of degree $n$ functions is an enormous collection of different maps, and heuristically, as long as the produced field qualitatively resembles the target field, they are likely to belong to the same homotopy class. Notice from Fig. \ref{fig:setup}c that although the fidelity of our synthesized field is limited by imperfections, especially of the SLM, the resulting Skyrmion number remains close to the target Skyrmion number. As an aside, the technique we have adopted here for Skyrmion generation is capable of producing Skyrmions of significantly higher order than the usual superposition of LG modes. 

Secondly, in all experiments, linear interpolation followed by a Gaussian filter and projection onto $S^2$ is applied in post-processing before the numerical evaluation of the Skyrmion number. This achieves two important goals. Firstly, smoothing helps limit the effects of noise on computations of partial derivatives, and secondly, smoothing homogenizes perturbations that have occurred in the boundary, therefore allowing for compactification. The latter is true because, by design, both our field and aberrations have been chosen to be relatively uniform for $r>1$. Note also that our post-processing method is permissible from the perspective of topology as Gaussian filtering---provided the standard deviation of the kernel is not too large as to cause points to cross hemispheres---can be regarded as a homotopy of the field by linearly interpolating standard deviation from 0.

Lastly, Fig. \ref{fig:results} shows the error defined by 
\begin{equation}
\label{eq:experiment_error}
    \text{error} = \text{Computed Skyrmion number} - \text{Target Skyrmion number}
\end{equation}
across different target Skyrmion numbers and experiment configurations. While the computed error is reasonably small, notice that there is a general trend for error to increase as Skyrmion number increases. One possible explanation is that numerical errors in differentiation and integration typically scale with the sup norm of suitable partial derivatives. Therefore, increasing the complexity of the field necessarily impacts the accuracy that can be achieved in numerical integration. Another possibility is that due to the limitations of the SLMs, such as discontinuities along pixels and limited resolution, the topological properties of Skyrmions of high order are less stable and, therefore, more affected by aberrations. 

Despite possible difficulties of resolving Skyrmions with a large Skyrmion number, our experiments show that for degrees up to 10, the computed Skyrmion number from polarimetric measurements---even in the presence of aberrating elements---remains close to the target Skyrmion number. With a maximum error of about 0.1, our results demonstrate the robustness of the topology of optical Skyrmions and support the use of Skyrmions in high-density data applications such as optical communications and photonic computing.  

Lastly, note that there is interesting scope for investigating the limit for which topological protection fails to hold, but due to the already lengthy nature of this work, we defer such an investigation to a future paper.

\section{Discussion}

First and foremost, we would like to emphasize that there exist complex spatially varying media for which topological protection does not hold. A direct example comes from specially designed gradient index systems \cite{Shen2023}, where one may explicitly show this to be true (see Supplementary Note 5). The reason this system does not offer topological protection is that its material parameters are not compactifiable. As a result, an incident Skyrmion loses its compactifiable structure as it propagates through the medium, and even if the resultant field is a Skyrmion, propagation through the system does not descend onto to homotopy in the sense of Methods 3.

In this paper, we do not proclaim that the optical Skyrmion is resilient to all perturbations; rather, we have tried to develop a definition of the Skyrmion and a corresponding framework for establishing results of topological robustness that can be derived from this definition. There is, however, significant scope to expand on these results. Whether or not, and to what extent, topological protection holds remains an important question in the study of optical Skyrmions, and both positive and negative results help elucidate fundamental properties of the optical Skyrmion. Moreover, negative results may even be helpful in the context of computing, where manipulation of the Skyrmion number can be used to perform integer arithmetic. 

We would also like to stress that the ability to generate Skyrmions of such high order is peculiar to the optical setting. Unlike magnetic Skyrmions, which are topological solitons in a true sense and have a notion of energy stability, such a concept has no analog in the optical domain. This energy stability in magnetic Skyrmions gives rise to a physical interpretation of topological protection beyond the abstract mathematical one stated in this paper, namely energy input is necessary to transition from one stable topological state to another. However the only interpretation accessible to the optical Skyrmion is the mathematical one. While this lack of an energy barrier has implications for the topological stability of optical Skyrmions, it also has significant advantages in terms of data density, where energy constraints do not impede the formation of high-order Skyrmions as we demonstrate in our experiments. In this sense, the theory of optical Skyrmions is not so much the study of a naturally topological system but rather about the ability to embed topological character into electromagnetic fields through techniques of structured light and about investigating the limits of this topological character. 

Lastly, we consider here the significance of the assumptions of smoothness and compactifiability we have made in establishing our theory. Note firstly that smoothness is a condition we introduced to ease our discussion of the topic and can, in fact, be greatly weakened. Indeed, most of what we have done holds directly for $C^1$ functions, but we can even generalize to continuous functions by swapping out the De Rham cohomology for singular cohomology. In this case, the integral equation for the Skyrmion number may not be well-defined as partial derivatives need not necessarily exist, but general density results guarantee that a continuous $\mathcal{S}$ is, in a suitable sense, close to some smooth $\mathcal{S}'$ of equivalent degree for which the integral equation holds. Importantly, as continuous functions are a much larger class than smooth functions, being able to handle continuous maps grants greater applicability of our theory to the real world.

Compactifiability, on the other hand, is nonoptional, and while many vector beams naturally satisfy this condition, it is still of practical and theoretical interest to extend the notion of the Skyrmion to a broader class of functions. Here, we propose one such way. 

The most obvious way of generalizing the Skyrmion is to simply drop the compactifiable condition and declare
\begin{equation}
\label{eq: Fractional Skyrmion Number}
    \text{the Skyrmion Number of $\mathcal{S}$} \equiv \text{SkyN}(\mathcal{S}) = \int_U \mathcal{S}^\ast \omega_0 = \frac{1}{4\pi}\iint_U S \cdot \left(\frac{\partial S}{\partial x} \times \frac{\partial S}{\partial y}\right) dxdy.
\end{equation}
In doing so, we gain the ability to assign arbitrary fields with a Skyrmion number but lose two important properties. Firstly, $\text{SkyN}(\mathcal{S})$ is no longer guaranteed to be an integer, and secondly, $\text{SkyN}(\mathcal{S})$ can no longer be interpreted as the degree of a smooth map. Therefore, the Skyrmion number, as defined above, is not a topological number in the true sense, and the cost of this generality is that we muddy the topological interpretation of the Skyrmion. Nonetheless, we can still proceed as in Methods 8. 

We call Skyrmions defined in this more general way ``fractional Skyrmions'' and discuss the manners in which their topological nature can be interpreted in Methods 9. Note that while other works have also brought up notions of a fractional Skyrmion, the interpretations we give in Methods 9 prescribe these fields a precise topological character and, importantly, a notion of homotopy invariance that can be used to establish conditions, albeit weaker, for the Skyrmion number integral to remain unchanged. In the case of optical fields, we can, as before, give sufficient conditions for the preservation of the Skyrmion number through spatially varying retarders, diattenuators, depolarizing media, and their combinations. In Methods 10, we flesh out the mathematical details of this argument. Lastly, notice that the abstract results we prove here are directly applicable to important real-world situations where the boundary conditions of the field are sufficiently constrained, such as propagation in a conducting waveguide. 

To conclude, in this paper we proposed a topological definition of the Skyrmion that clarifies conditions for the Skyrmion number integral to evaluate to an integer, namely compactifiability, and made clear the properties of the underlying physics that allows such topological structures to propagate in electromagnetic fields, namely the ellipticity of the Helmholtz equation. Using our new definition, we establish general procedures for handling perturbations to Skyrmion fields, and demonstrated, both theoretically and experimentally, the robustness of an optical Skyrmion's topological structure against important polarization aberrations. 

As a concluding remark, we emphasize once again that the formalism of compactification implies the existence of $S^2$-valued fields, not uniform on the boundary, for which an integer topological number can be defined. While the prevailing approach to optical Skyrmions has been to consider fields constant on their boundaries, there is certainly room to explore topological textures with more complicated symmetries.

We believe the work presented in this paper establishes a solid foundation for further investigations into the topological protection of optical Skyrmions and hope our findings can spur further developments in this exciting field.

\newpage

\section*{Acknowledgement}
The authors would like to thank the support of St John’s College and the John Fell Fund (University of Oxford) (C.H.); the European Research Council (AdOMiS, no. 695140) (C.H. and M.J.B.); Shenzhen Key Fundamental Research Project (No. JCYJ20210324120012035) (H.H.). The authors would also like to thank Prof.\ Andrew Forbes of the University of the Witwatersrand for his valuable comments.

\section*{Competing Interests}
The authors declare no competing interests.

\section*{Additional Information}
Correspondence and requests for materials should be addressed to A.W. or C.H.

\newpage
\renewcommand{\figurename}{Extended Fig.}
\setcounter{figure}{0}

\clearpage
\bibliographystyle{unsrt}
\bibliography{main}

\begin{thebibliography}{10}

\bibitem{Nagaosa2013}
N.~Nagaosa and Y.~Tokura.
\newblock Topological properties and dynamics of magnetic skyrmions.
\newblock {\em Nature Nanotechnology}, 8(12):899–911, 2013.

\bibitem{yu_real-space_2010}
X.~Z. Yu, Y.~Onose, N.~Kanazawa, J.~H. Park, J.~H. Han, Y.~Matsui, N.~Nagaosa, and Y.~Tokura.
\newblock Real-space observation of a two-dimensional skyrmion crystal.
\newblock {\em Nature}, 465(7300):901--904, 2010.

\bibitem{shi_spin_2021}
P.~Shi, L.~Du, and X.~Yuan.
\newblock Spin photonics: from transverse spin to photonic skyrmions.
\newblock {\em Nanophotonics}, 10(16):3927--3943, 2021.

\bibitem{Fert2017}
A.~Fert, N.~Reyren, and V.~Cros.
\newblock Magnetic skyrmions: Advances in physics and potential applications.
\newblock {\em Nature Reviews Materials}, 2(7), 2017.

\bibitem{cortes-ortuno_thermal_2017}
D.~Cortés-Ortuño, W.~Wang, M.~Beg, R.~A. Pepper, M.~Bisotti, R.~Carey, M.~Vousden, T.~Kluyver, O.~Hovorka, and H.~Fangohr.
\newblock Thermal stability and topological protection of skyrmions in nanotracks.
\newblock {\em Scientific Reports}, 7(1):4060, 2017.

\bibitem{je_direct_2020}
S.~Je, H.~Han, S.~K. Kim, S.~A. Montoya, W.~Chao, I.~Hong, E.~E. Fullerton, K.~Lee, K.~Lee, M.~Im, and J.~Hong.
\newblock Direct {Demonstration} of {Topological} {Stability} of {Magnetic} {Skyrmions} via {Topology} {Manipulation}.
\newblock {\em ACS Nano}, 14(3):3251--3258, 2020.

\bibitem{Kang2016}
W.~Kang, Y.~Huang, C.~Zheng, W.~Lv, N.~Lei, Y.~Zhang, X.~Zhang, Y.~Zhou, and W.~Zhao.
\newblock Voltage controlled magnetic skyrmion motion for racetrack memory.
\newblock {\em Scientific Reports}, 6(1), 2016.

\bibitem{song_skyrmion-based_2020}
K.~M. Song, J.~Jeong, B.~Pan, X.~Zhang, J.~Xia, S.~Cha, T.~Park, K.~Kim, S.~Finizio, J.~Raabe, J.~Chang, Y.~Zhou, W.~Zhao, W.~Kang, H.~Ju, and S.~Woo.
\newblock Skyrmion-based artificial synapses for neuromorphic computing.
\newblock {\em Nature Electronics}, 3(3):148--155, 2020.

\bibitem{Zhang2015}
X.~Zhang, M.~Ezawa, and Y.~Zhou.
\newblock Magnetic skyrmion logic gates: Conversion, duplication and merging of skyrmions.
\newblock {\em Scientific Reports}, 5(1), 2015.

\bibitem{Tsesses2018}
S.~Tsesses, E.~Ostrovsky, K.~Cohen, B.~Gjonaj, N.~H. Lindner, and G.~Bartal.
\newblock Optical skyrmion lattice in evanescent electromagnetic fields.
\newblock {\em Science}, 361(6406):993–996, 2018.

\bibitem{Gao2020}
S.~Gao, F.~C. Speirits, F.~Castellucci, S.~Franke-Arnold, S.~M. Barnett, and J.~B. G\"otte.
\newblock Paraxial skyrmionic beams.
\newblock {\em Phys. Rev. A}, 102:053513, 2020.

\bibitem{du_deep-subwavelength_2019}
L.~Du, A.~Yang, A.~V. Zayats, and X.~Yuan.
\newblock Deep-subwavelength features of photonic skyrmions in a confined electromagnetic field with orbital angular momentum.
\newblock {\em Nature Physics}, 15(7):650--654, 2019.

\bibitem{Shen2022}
Y.~Shen, E.~C. Martínez, and C.~Rosales-Guzmán.
\newblock Generation of optical skyrmions with tunable topological textures.
\newblock {\em ACS Photonics}, 9(1):296–303, 2022.

\bibitem{He2022}
C.~He, Y.~Shen, and A.~Forbes.
\newblock Towards higher-dimensional structured light.
\newblock {\em Light: Science \& Applications}, 11(1), 2022.

\bibitem{shen_optical_2023}
Y.~Shen, Q.~Zhang, P.~Shi, L.~Du, X.~Yuan, and A.~V. Zayats.
\newblock Optical skyrmions and other topological quasiparticles of light.
\newblock {\em Nature Photonics}, 2023.

\bibitem{Bai2020}
C.~Bai, J.~Chen, Y.~Zhang, D.~Zhang, and Q.~Zhan.
\newblock Dynamic tailoring of an optical skyrmion lattice in surface plasmon polaritons.
\newblock {\em Optics Express}, 28(7):10320, 2020.

\bibitem{Lin2021}
W.~Lin, Y.~Ota, Y.~Arakawa, and S.~Iwamoto.
\newblock Microcavity-based generation of full poincar\'e beams with arbitrary skyrmion numbers.
\newblock {\em Phys. Rev. Res.}, 3:023055, 2021.

\bibitem{Shen2021_Super}
Y.~Shen, Y.~Hou, N.~Papasimakis, and N.~I. Zheludev.
\newblock Supertoroidal light pulses as electromagnetic skyrmions propagating in free space.
\newblock {\em Nature Communications}, 12(1), 2021.

\bibitem{Cisowski2023}
C.~Cisowski, C.~Ross, and S.~Franke-Arnold.
\newblock Building paraxial optical skyrmions using rational maps.
\newblock {\em Advanced Photonics Research}, 4(4), 2023.

\bibitem{Chao2019}
C.~He, J.~Chang, Q.~Hu, J.~Wang, J.~Antonello, H.~He, S.~Liu, J.~Lin, B.~Dai, D.~S. Elson, P.~Xi, H.~Ma, and M.~J. Booth.
\newblock Complex vectorial optics through gradient index lens cascades.
\newblock {\em Nature Communications}, 10(1), 2019.

\bibitem{shi_strong_2020}
P.~Shi, L.~Du, and X.~Yuan.
\newblock Strong spin–orbit interaction of photonic skyrmions at the general optical interface.
\newblock {\em Nanophotonics}, 9(15):4619--4628, 2020.

\bibitem{lei_photonic_2021}
X.~Lei, A.~Yang, P.~Shi, Z.~Xie, L.~Du, A.~V. Zayats, and X.~Yuan.
\newblock Photonic {Spin} {Lattices}: {Symmetry} {Constraints} for {Skyrmion} and {Meron} {Topologies}.
\newblock {\em Physical Review Letters}, 127(23):237403, 2021.

\bibitem{Liu2022}
C.~Liu, S.~Zhang, S.~A. Maier, and H.~Ren.
\newblock Disorder-induced topological state transition in the optical skyrmion family.
\newblock {\em Physical Review Letters}, 129(26), 2022.

\bibitem{ye2024theory}
Z.~Ye, S.~M. Barnett, S.~Franke-Arnold, J.~B. G\"{o}tte, A.~McWilliam, F.~C. Speirits, and C.~M. Cisowski.
\newblock Theory of paraxial optical skyrmions.
\newblock Preprint at \url{https://arxiv.org/abs/2404.11530}, 2024.

\bibitem{Shen2023}
Y.~Shen, C.~He, Z.~Song, B.~Chen, H.~He, Y.~Ma, J.~A.~J. Fells, S.~J. Elston, S.~M. Morris, M.~J. Booth, and A.~Forbes.
\newblock Topologically controlled multiskyrmions in photonic gradient-index lenses.
\newblock {\em Physical Review Applied}, 21(2):024025, 2024.

\bibitem{skyrme1961non}
T.~H.~R. Skyrme.
\newblock A non-linear field theory.
\newblock {\em Proceedings of the Royal Society of London. Series A. Mathematical and Physical Sciences}, 260(1300):127--138, 1961.

\bibitem{ornelas_non-local_2024}
P.~Ornelas, I.~Nape, R.~de~Mello~Koch, and A.~Forbes.
\newblock Non-local skyrmions as topologically resilient quantum entangled states of light.
\newblock {\em Nature Photonics}, 18(3):258--266, 2024.

\bibitem{ornelas2024topologicalrejectionnoisenonlocal}
P.~Ornelas, I.~Nape, R.~de~Mello~Koch, and A.~Forbes.
\newblock Topological rejection of noise by non-local quantum skyrmions.
\newblock Preprint at \url{https://arxiv.org/abs/2403.02031}, 2024.

\bibitem{he2023universal}
C.~He et~al.
\newblock A universal optical modulator for synthetic topologically tuneable structured matter.
\newblock Preprint at \url{https://arxiv.org/abs/2311.18148}, 2023.

\bibitem{teng_physical_2023}
H.~Teng, J.~Zhong, J.~Chen, X.~Lei, and Q.~Zhan.
\newblock Physical conversion and superposition of optical skyrmion topologies.
\newblock {\em Photonics Research}, 11(12):2042--2053, 2023.

\bibitem{kinsey_topology_2012}
L.~Christine Kinsey.
\newblock {\em Topology of {Surfaces}}.
\newblock Undergraduate {Texts} in {Mathematics}. Springer New York, New York, NY, 1st edition, 2012.

\bibitem{chipman_polarized_2018}
R.~Chipman, W.~S.~T. Lam, and G.~Young.
\newblock {\em Polarized {Light} and {Optical} {Systems}}.
\newblock CRC Press, Boca Raton, 1st edition, 2018.

\bibitem{mcguire_polarization_1994}
J.~P. McGuire and R.~A. Chipman.
\newblock Polarization aberrations. 1. {Rotationally} symmetric optical systems.
\newblock {\em Applied Optics}, 33(22):5080--5100, 1994.

\bibitem{he_vectorial_2023}
C.~He, J.~Antonello, and M.~J. Booth.
\newblock Vectorial adaptive optics.
\newblock {\em eLight}, 3(1):23, 2023.

\bibitem{he_polarisation_2021}
C.~He, H.~He, J.~Chang, B.~Chen, H.~Ma, and M.~J. Booth.
\newblock Polarisation optics for biomedical and clinical applications: a review.
\newblock {\em Light: Science \& Applications}, 10(1):194, 2021.

\bibitem{hu_arbitrary_2020}
Q.~Hu, Y.~Dai, C.~He, and M.~J. Booth.
\newblock Arbitrary vectorial state conversion using liquid crystal spatial light modulators.
\newblock {\em Optics Communications}, 459:125028, 2020.

\end{thebibliography}


\begin{thebibliography}{1}

\bibitem{naber_topology_2011}
Gregory~L Naber.
\newblock {\em Topology, {Geometry} and {Gauge} fields {Interactions}}.
\newblock Applied {Mathematical} {Sciences}. Springer, New York, NY, 2nd edition, 2011.

\bibitem{lu_homogeneous_1994}
S.~Lu and R.~A. Chipman.
\newblock Homogeneous and inhomogeneous {Jones} matrices.
\newblock {\em Journal of the Optical Society of America A}, 11(2):766--773, 1994.

\bibitem{Jorge2022}
G.~P.~J. Jorge and R.~Ossikovski.
\newblock {\em Nondepolarizing Media}, page 125–158.
\newblock CRC Press, 2022.

\bibitem{Ossikovski09}
R.~Ossikovski.
\newblock Analysis of depolarizing mueller matrices through a symmetric decomposition.
\newblock {\em J. Opt. Soc. Am. A}, 26(5):1109--1118, 2009.

\end{thebibliography}

\newpage

\section*{Methods}
\setcounter{section}{0}

\section{Defining the Skyrmion}

The central tool in understanding the topology of the Skyrmion is the De Rham cohomology \citesupp{naber_topology_2011}, which is the mathematical theory that allows one to prescribe spatially varying fields taking values in $S^2$, such as polarization fields, topological character via an integral equation. It is a well-known result in multivariable calculus that on a contractible open subset of $\mathbb{R}^n$, if a $k$-form $\omega$ satisfies $d\omega =0$, then there necessarily exists a $(k-1)$-form $\eta$ so that $d\eta = \omega$. This can be understood as a generalization of the more commonly seen facts in physics that a divergence-free vector field can be written as the curl of a vector potential, and a curl-free vector field can be written as the gradient of a scalar potential. Therefore, if there exists a $k$-form $\omega$ satisfying $d\omega =0$ but no corresponding $(k-1)$-form $\eta$ so that $d\eta = \omega$, this gives us topological information about the underlying space, namely that it cannot be contractible. The De Rham cohomology expands this idea by introducing a collection of algebraic objects that can be assigned to any smooth manifold $X$ called its De Rham cohomology groups $H_{\text{de R}}^k(X)$ and whose algebraic structure subtly captures information about how the underlying space fails to be contractible. 

Given a compact, connected, oriented, smooth, $n$-dimensional manifold $X$, one can show using Stokes' theorem that the top cohomology group is isomorphic to $\mathbb{R}$ as a vector space via the map $\omega \mapsto \int_X \omega$. Given two such manifolds $X$ and $Y$, and a smooth map $f\colon X\longrightarrow Y$ between them, the naturally induced map in cohomology given by pullback is then a $\mathbb{R}$-linear map $f^\# \colon H_{\text{de R}}^{n}(Y) \cong \mathbb{R} \longrightarrow H_{\text{de R}}^{n}(X) \cong \mathbb{R}$. As the only $\mathbb{R}$-linear map from $\mathbb{R}$ to itself is of the form $x \mapsto cx$ for some $c\in \mathbb{R}$, we may assign to every such $f$ a real number $\deg f \coloneqq c$. Using an argument invoking the regular value theorem, one can further show that the degree of any map takes values in the integers. The homotopy invariance of the degree is then the key mathematical concept that captures our intuitions of what topological robustness to perturbations should be.

With the above machinery on hand, we are in the position to define the Skyrmion. Given a field that takes values in the 2-sphere, checking against the hypotheses in the definition of the degree, we see that the target space of the field is already a compact, connected, oriented, smooth, $2$-dimensional manifold. The domain of the field is, however, not naturally of this form. Therefore, extra work is required to ``non-trivialize'' the domain of the field. If we are interested in fields that occupy three dimensions, there is an immediate solution by restricting to a submanifold of $\mathbb{R}^3$ with the required properties. Defined like this, the Skyrmion can be regarded, topologically, as a strict higher dimensional generalization of orbital angular momentum, which derives its topological character in exactly this way, except by restricting phase to a suitable one-dimensional compact and connected submanifold of $\mathbb{R}^2$. 

However, the drawback of using the above approach is that the Skyrmion becomes a fundamentally three-dimensional object rather than having the more commonly understood planar structure. In the context of optical fields, such an interpretation of the Skyrmion also loses the natural notion of propagation, and we would much rather define the Skyrmion by considering polarization fields when restricted to the transverse plane. Here, the mathematics inevitably becomes more technical. Instead of simply restricting to a submanifold with the necessary properties, one must extend the trivial two-dimensional plane into a non-trivial manifold. However, this process, which we formalize in Methods 2, restricts the fields that can and cannot be considered a Skyrmion. We call fields compatible with the extension of its underlying domain compactifiable and note that the boundary conditions of the field determine whether compactification holds. For instance, a field that approaches a constant value on its boundary can be compactified through stereographic projection, while a periodic field can be compactified by projection onto the torus  $\mathbb{R}^2/\mathbb{Z}^2 \cong \mathbb{T}^2$.

The significance of what are ostensibly just mathematical technicalities is that they clarify precisely what properties of the field allow for the definition of an integer-valued topological number, which, as mentioned in the introduction, is the lynchpin in establishing topological robustness. Moreover, from our definition, we can show that any compactifiable $\mathcal{S}\colon U \longrightarrow S^2$ satisfies an uncountable collection of non-trivial integral equations given by 
\begin{equation}
    \int_U \mathcal{S}^\ast \omega = \deg \mathcal{S}\int_{S^2} \omega
\end{equation}
for every $\omega \in \Lambda^2(S^2)$, of which the usual integral equation is just a special case (see Supplementary Note 1). In fact, satisfying the usual integral equation does not capture any topological structure of the underlying field, and it is easy to construct a field where the usual integral equation evaluates to an integer but not an uncountable number of other integral equations that we expect a truly topological field to satisfy (take, for example, a meron of second-order vorticity). In this sense, defining the Skyrmion through the integral equation is simply nonsensical from the perspective of topology. 

\section{Compactifiability}

Suppose we have an field $\mathcal{S}\colon U \longrightarrow S^2$ defined on an open subset $U$ of $\mathbb{R}^2$. If there exists a compact, connected, oriented 2-dimensional manifold $X$ and a smooth map $\psi\colon U \longrightarrow X$ satisfying the following properties:
\begin{enumerate}
    \item $\psi$ is an orientation preserving diffeomorphism of $U$ onto its image,
    \item $\psi(U)$ is a dense subset of $X$ with full measure,
    \item $\mathcal{S}\circ \psi^{-1}$ extends via continuity to a smooth map $\tilde{\mathcal{S}}$ on all of $X$,
\end{enumerate}
we call $\mathcal{S}$ compactifiable, $\tilde{\mathcal{S}}$ its compactification, and prescribe the field  $\mathcal{S}$ with a Skyrmion number $\deg \mathcal{S} \coloneqq \deg \tilde{\mathcal{S}}$, where we understand $\deg \tilde{\mathcal{S}} \in \mathbb{Z}$ as the degree of a smooth map originating from De Rham cohomology. The topological nature of the Skyrmion then results from the fact that two Skyrmion fields $\tilde{\mathcal{S}}$ and $\tilde{\mathcal{S}}'$ have the same degree if they are smoothly homotopic. Note that for fields with compact domains, a more natural way of compactification is via quotients. However, such an approach is not applicable to $\mathbb{R}^2$, which is the most natural domain to consider in many cases. In fact, there is no loss of generality in adopting the notion of compactification presented here as a field defined on a compact set that descends onto a surface through a suitable quotient can also be compactified, in the sense we have defined above, by considering the restriction of the field onto the interior of its domain. 

To derive the usual Skyrmion number integral from this definition, let $\iota \colon S^2 \hookrightarrow \mathbb{R}^3$ be the inclusion, $\omega_0 \in \Lambda^2(S^2)$ the normalized volume form $\omega_0 = \frac{1}{4\pi}\iota^\ast(x^1 dx^2 \wedge dx^3 - x^2 dx^2 \wedge dx^3 + x^3 dx^1\wedge dx^2)$ where $x^1, x^2$ and $x^3$ are the standard coordinate functions on $\mathbb{R}^3$, and $S \coloneqq \iota \circ \mathcal{S}$. If $x$ and $y$ are the standard coordinate functions on $U$, then
\begin{align}
    \deg \mathcal{S} \coloneqq \int_X \tilde{\mathcal{S}}^\ast \omega_0 = \int_{\psi(U)} \tilde{\mathcal{S}}^\ast \omega_0 & = \int_U \psi^\ast(\tilde{\mathcal{S}}^\ast\omega_0) \nonumber \\ & = \int_U (\tilde{\mathcal{S}}\circ \psi)^\ast \omega_0 = \int_U \mathcal{S}^\ast\omega_0 = \frac{1}{4\pi} \iint_U S\cdot \left(\frac{\partial S}{\partial x} \times \frac{\partial S}{\partial y}\right) dxdy, 
\end{align}
which is the Skyrmion number integral in its usual guise (see Supplementary Note 1). Note that the technical conditions (1) and (2) are precisely the conditions needed to pullback $\omega$ from $S^2$ to $U$, with the density of $\psi(U)$ giving the first equality above and $\psi$ being an orientation preserving diffeomorphism giving the second. Notice also that although we required the existence of $\psi$ for compactification, the Skyrmion number itself is independent of the choice of $\psi$. The Skyrmion number is, therefore, a genuine characteristic of the field. 

\section{A Homotopy Approach to Topological Protection}

Using the notion of compactifiability, we can determine the effects of any perturbation on the topology of a Skyrmion using the following simple procedure. 

\begin{enumerate}
    \item Write the perturbed field $\mathcal{S}'\colon U \longrightarrow S^2$ in the form $\mathcal{S}'(p)=f(\mathcal{S}(p), \alpha_1(p), \ldots, \alpha_k(p))$ for some smooth function $f$ and smoothly varying parameters $\alpha_i$, $i=1, \ldots, k$, that characterize the perturbation. 
    \item Construct a smooth homotopy from $\tilde{\mathcal{S}}'$ to a map of the form $\bar{f} \circ \tilde{\mathcal{S}}$ for some smooth $\bar{f}$. Then, by the smooth homotopy invariance of the degree, we have $\deg\mathcal{S}' \coloneqq \deg\tilde{\mathcal{S}}' = \deg(\bar{f}\circ \tilde{\mathcal{S}}) = (\deg\bar{f})(\deg\tilde{\mathcal{S}}) \eqqcolon (\deg\bar{f}) (\deg\mathcal{S})$.
\end{enumerate}

The key benefit of this approach is that we can often reduce our analysis from the possibly intractable Skyrmion number integral to the much simpler computation $\deg \bar{f} = \sum_{i=1}^k \text{sign}(\bar{f}, p_i)$, where $\{p_1, \ldots, p_k\} = \bar{f}^{-1}(q)$ for any regular value $q$ of $\bar{f}$, $\text{sign}(\bar{f}, p_i) = 1$ if $\bar{f}_{\ast p_i}$ is orientation preserving and $\text{sign}(\bar{f}, p_i) = -1$ if $\bar{f}_{\ast p_i}$ is orientation reversing. As a technical aside, observe that in order for $\mathcal{S}'$ to be a Skyrmion field, it is sufficient for $\alpha_i\circ \psi^{-1}$ to extend via continuity to a smooth map $\tilde{\alpha}_i$ on all of $X$. In which case, the smoothness of $f$ implies that $\mathcal{S}'\circ\psi^{-1}$ also extends via continuity to the smooth map $\tilde{\mathcal{S}}'\colon X \rightarrow S^2$, $\tilde{\mathcal{S}}'(x)=f(\tilde{\mathcal{S}}(x), \tilde{\alpha}_1(x), \ldots, \tilde{\alpha}_k(x))$. Lastly, as mentioned in the main text, if the codomain of $\psi$ is $S^2$, then the Hopf degree theorem guarantees that topological protection holds if and only if the above procedure holds for some homotopy. 

\section{Spatially Varying Retarders}
Consider a spatially varying retarder characterized by its Jones matrix $e^{i\phi(p)}J(p)$, where $J\colon U \longrightarrow SU(2)$ represents the relative phase shift between fast and slow axes, and $\phi \colon U \longrightarrow \mathbb{R}$ the absolute phase \citesupp{lu_homogeneous_1994}. Through our proposed methodology, we show that topological protection is guaranteed whenever $J$ is smooth and compactifiable in a compatible way with the incident field, that is $J\circ \psi^{-1}$ extends via continuity to a smooth map for the same $\psi$ that compactifies $\mathcal{S}$. We call compactifiability in this compatible way the duality condition for guaranteeing topological protection. Adopting the language introduced in Methods 3, we first write 
\begin{equation}
    \mathcal{S}'(p) = f(\mathcal{S}(p), J(p)) = A^T\text{Spin}(J(p))A\mathcal{S}(p) 
\end{equation}
where $\text{Spin}\colon SU(2) \longrightarrow SO(3)$ is the usual Spin map 
\begin{equation}
    \begin{pmatrix}
        a+bi & c+di \\ -c+di & a-bi
    \end{pmatrix} \mapsto \begin{pmatrix}
        a^2-b^2-c^2+d^2 & 2ab+2cd & -2ac+2bd \\ -2ab+2cd & a^2-b^2+c^2-d^2 & 2ad+2bc \\ 2ac+2bd & 2bc-2ad & a^2+b^2-c^2-d^2
    \end{pmatrix}
\end{equation}
and
\begin{equation}
    A = \begin{pmatrix}
        0 & 1 & 0 \\ 0 & 0 & 1 \\ 1 & 0 & 0
    \end{pmatrix}.
\end{equation}

To construct a suitable homotopy, let $\Phi\colon SU(2) \longrightarrow S^3 \subseteq \mathbb{R}^4$ be the canonical diffeomorphism given by 
\begin{equation}
    \Phi\begin{pmatrix}
    a+bi & c+di \\ -c+di & a-bi
    \end{pmatrix} = \begin{pmatrix}
        a \\ b \\c \\ d
    \end{pmatrix}.
\end{equation}
 As $X$ is a 2-dimensional manifold and $S^3$ a 3-dimensional manifold, $\Phi \circ \tilde{J} \colon X \longrightarrow S^3$ is never surjective. There therefore exists some $q\in S^3$ not in the image of $\Phi \circ \tilde{J}$. We now define a continuous homotopy from $\Phi \circ \tilde{J}$ to the constant map taking value $-q$ as follows. Let $H\colon X\times[0,1] \longrightarrow S^3$ be 
\begin{equation}
    H(x,t) = \frac{(1-t)\Phi(\tilde{J}(x))-tq}{\lVert (1-t)\Phi(\tilde{J}(x))-tq \rVert}.
\end{equation}
Since $\Phi(\tilde{J}(x)) \neq q$, $\lVert (1-t)\Phi(\tilde{J}(x))-tq \rVert$ is never zero, and hence $H$ is well-defined. Moreover, $H$ is easily seen to be continuous, and satisfies $H(x,0) = \Phi(\tilde{J}(x))$ and $H(x,1) = -q$, as required. By a further application of the Whitney Approximation Theorem (see Supplementary Note 2), we can find a smooth homotopy $\tilde{H}\colon X \times (-\epsilon, 1+\epsilon)\longrightarrow S^3$ with $\tilde{H}(x,0) = \Phi(\tilde{J}(x))$ and $\tilde{H}(x,1) = -q$. Using this, we may smoothly homotope $\tilde{\mathcal{S}}'$ to $A^T\text{Spin}(\Phi^{-1}(-q))A \tilde{\mathcal{S}}$ via $F\colon X \times (-\epsilon, 1+\epsilon) \longrightarrow S^2$,
\begin{equation}
    F(x, t) = A^T\text{Spin}(\Phi^{-1}(\tilde{H}(x,t)))A\tilde{\mathcal{S}}(x)
\end{equation}
Then, by the smooth homotopy invariance of $\deg$, we have
\begin{equation}
    \deg \mathcal{S}' \coloneqq \deg\tilde{\mathcal{S}}' = \deg (A^T\text{Spin}(\Phi^{-1}(-q))A\tilde{\mathcal{S}}). 
\end{equation}
Notice now that $A^T\text{Spin}(\Phi^{-1}(-q))A\tilde{\mathcal{S}}$ is just $\bar{f} \circ \tilde{\mathcal{S}}$ where $\bar{f} \colon S^2 \longrightarrow S^2$ is a rotation. Since every rotation is an orientation preserving diffeomorphism, $\deg \bar{f} = 1$. We therefore have
\begin{equation}
    \deg (A^T\text{Spin}(\Phi^{-1}(-q))A\tilde{\mathcal{S}}) = \deg (\bar{f}\circ \tilde{\mathcal{S}}) = (\deg \bar{f})(\deg \tilde{\mathcal{S}}) = \deg \tilde{\mathcal{S}} \eqqcolon \deg \mathcal{S},
\end{equation}
as required.

To support our mathematical proof, we numerically computed the Skyrmion number of an incident Skyrmion field after it passes through a complex spatially varying retarder. For convenience, we consider a special class of retarders given by the parametrization 
\begin{equation}
\label{eq: retarder_jones}
    J = \begin{pmatrix}
        \cos^2(\alpha)e^{i\Delta/2}+\sin^2(\alpha)e^{-i\Delta/2} & 2i\cos(\alpha)\sin(\alpha)\sin(\Delta/2)e^{-i\delta} \\
        2i\cos(\alpha)\sin(\alpha)\sin(\Delta/2)e^{i\delta} & \sin^2(\alpha)e^{i\Delta/2}+\cos^2(\alpha)e^{-i\Delta/2}
    \end{pmatrix}
\end{equation}
where $\alpha, \delta \colon U \longrightarrow \mathbb{R}$ are smooth maps defining a smoothly varying fast axis $(\cos(\alpha)e^{-i\delta/2}, \sin(\alpha)e^{i\delta/2})$, and $\Delta \colon U \longrightarrow \mathbb{R}$ the corresponding smoothly varying retardance. Note that spatial dependence has been suppressed here for clarity. 

\begin{figure}[!p]
    \centering
    \includegraphics[width=0.97\textwidth]{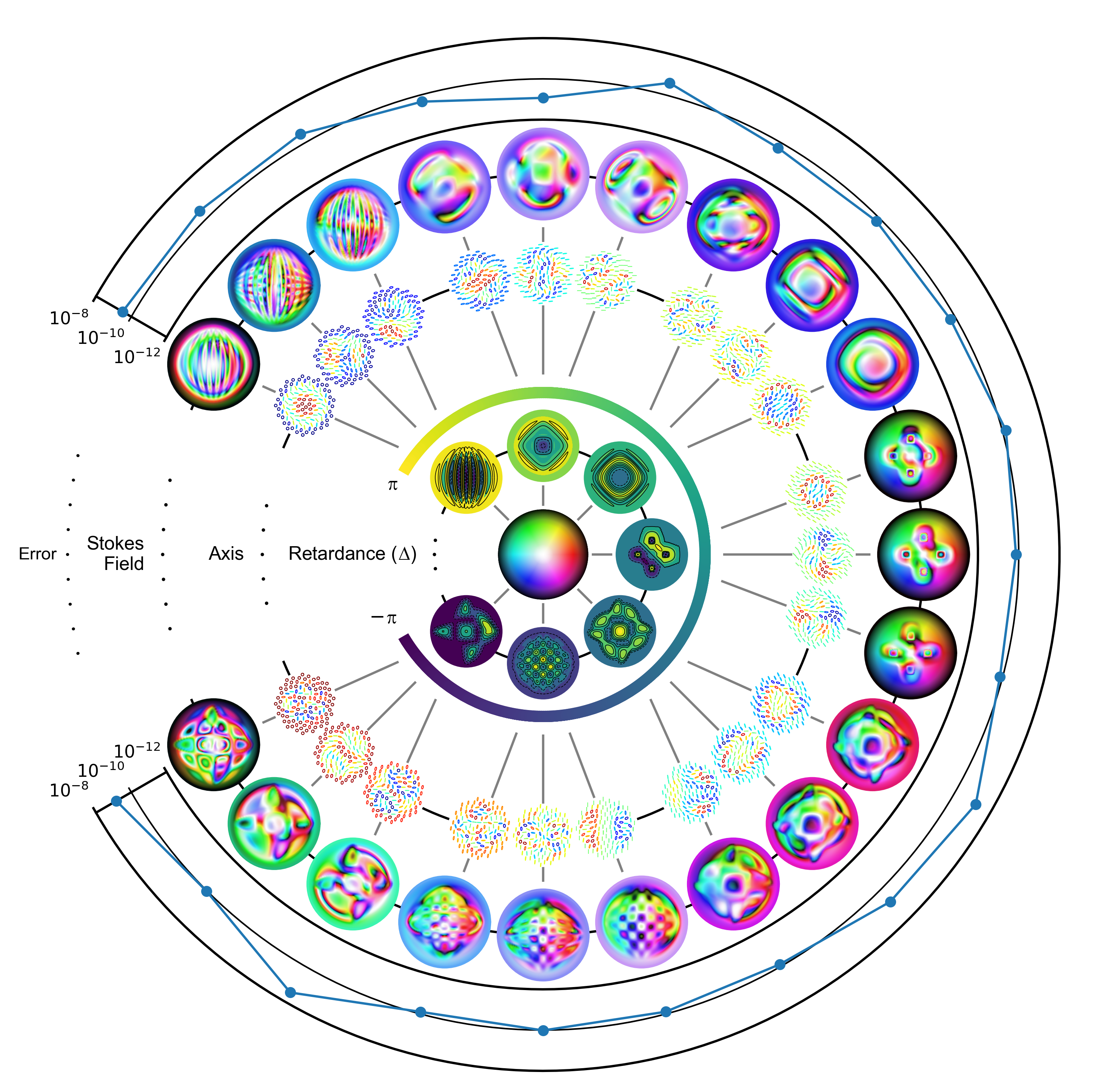}
    \caption{{\bf Skyrmions through spatially varying retarders.} The output Stokes field of a first-order N\'{e}el-type Skyrmion after passing through a retarder with spatially varying axis and retardance. The overall Jones matrix is given by equation (\ref{eq: retarder_jones}), with the distribution of $\Delta$ given by contour plots in the innermost ring and the axis distribution represented by ellipses in the second innermost ring. The output Stokes field is shown in the second outermost ring and the output error as defined in equation (\ref{eq: error}) is shown in the outermost ring. The Skyrmion numbers are computed using a central difference for partial derivatives and the `nquad' function from SciPy for integration. Note that machine epsilon for double-precision floating-points is of the other $10^{-16}$.}
    \label{fig:mixed_retarder}
\end{figure}

In all simulations, we take $U$ to be the open ball of radius $\pi/2$ centered at the origin, $\psi = \psi_2 \circ \psi_1$ the composition of $\psi_1(r\cos\theta, r\sin\theta) = (\tan(r)\cos\theta, \tan(r)\sin\theta)$ with the inverse stereographic projection from the south pole,
\begin{equation}
    \psi_2(x,y) = \frac{1}{1+x^2+y^2}\left(2x, 2y, 1-x^2-y^2\right),
\end{equation}
and standard N\'{e}el-type Skyrmions (see Supplementary Note 3) as inputs. A central difference is used to compute the partial derivatives of the output Skyrmion field, and an adaptive quadrature rule is used for numerical integration (the `nquad' function from the SciPy library). Extended Fig. \ref{fig:mixed_retarder} shows the output Stokes field, retarder axis, and retardance used in simulations. A relative error defined by 
\begin{equation}
\label{eq: error}
    \text{error} = \left\lvert \frac{\text{Computed Skyrmion Number} - \text{Initial Skyrmion Number}}{\text{Initial Skyrmion Number}}\right\rvert
\end{equation}
is also shown. From the figure, we see that numerical error lies in the range $10^{-8}$ to $10^{-12}$, which is strong evidence for the validity of our results.

\newpage

\section{Spatially Varying Diattenuators}

Consider a spatially varying horizontal linear diattenuator given by the Mueller matrix \citesupp{Jorge2022}
\begin{equation}
D = \frac{p_1^2}{1+\cos\kappa} \begin{pmatrix}
    1 & \cos \kappa & 0 & 0 \\ \cos \kappa & 1 & 0 & 0 \\ 0 & 0 & \sin \kappa & 0 \\ 0 & 0 & 0 & \sin \kappa
\end{pmatrix},
\end{equation}
where $\cos \kappa = (p_1^2-p_2^2)/(p_1^2+p_2^2)$ is the diattenuation, $p_1 \geq p_2 \geq 0$ the principal amplitude coefficients, and spatial dependence suppressed for clarity. Note that, unlike a retarder, the output polarization state through a diattenuator depends on the degree of polarization of the input. If we let $P\colon U \longrightarrow [0,1]$ be the spatially varying incident degree of polarization and take $\kappa \colon U \longrightarrow [0,\pi/2]$, our approach shows that provided $P$ and $\kappa$ are smooth, compactifiable in a compatible way with the incident field (duality condition), and satisfy $\cos \tilde{\kappa}(x) < \tilde{P}(x)$ for all $x \in X$, the Skyrmion number of the field is preserved.

To see why this is true, note first that we may write 
\begin{equation}
    \mathcal{S}' = f(\mathcal{S}, P, \kappa) = \frac{1}{L(\mathcal{S}, P, \kappa)}\begin{pmatrix}
            \cos\kappa + Ps_1\\ Ps_2\sin\kappa \\ Ps_3\sin\kappa
        \end{pmatrix} \\
\end{equation}
where
\begin{equation}
    L(\mathcal{S}, P, \kappa) = \sqrt{(1+Ps_1\cos\kappa)^2 + (P^2-1)\sin^2\kappa}.
\end{equation}
One important technicality is that for the output polarization state to be well-defined, we require $L(\mathcal{S},P,\kappa) \neq 0$. Noting that $0\leq P, \cos\kappa \leq 1$, it is clear that for fixed $P$ and $\kappa$, $(1+Ps_1\cos\kappa)^2 + (P^2-1)\sin^2\kappa$ is minimum when $s_1=-1$. In which case, $L^2 = (P-\cos\kappa)^2$. Therefore, $f(\mathcal{S}, P, \kappa)$ is well-defined provided $P\neq \cos \kappa$. 

Now, let $\bar{f} \colon S^2 \longrightarrow S^2$ be the map
\begin{equation}
    \bar{f}(s) = f(s, 1, \pi/2) = s.
\end{equation}
If $\cos \tilde{\kappa}(x) < \tilde{P}(x)$ for all $x \in X$, we may construct a continuous homotopy $H\colon X \times [0,1] \longrightarrow S^2$ from $\tilde{\mathcal{S}}'$ to $\bar{f}\circ\tilde{\mathcal{S}}$ by 
\begin{equation}
    H(x, t) = f(\tilde{\mathcal{S}}(x), (1-t)\tilde{P}(x) + t, (1-t)\tilde{\kappa}(x) + \pi t/2),
\end{equation}
where this is well-defined because $\cos((1-t)\tilde{\kappa}(x) + \pi t/2) \leq \cos\tilde{\kappa}(x) < \tilde{P}(x) \leq (1-t)\tilde{P}(x) + t$ for all $(x,t) \in X \times [0,1]$. By further application of the Whitney Approximation Theorem, we also have that $\tilde{\mathcal{S}}'$ and $\bar{f} \circ \tilde{\mathcal{S}}$ are smoothly homotopic. Then
\begin{equation}
    \deg \mathcal{S}'  \coloneqq \deg \tilde{\mathcal{S}}' = \deg(\bar{f} \circ \tilde{\mathcal{S}}) = \deg(\bar{f})\deg(\tilde{\mathcal{S}}) = \deg(\tilde{\mathcal{S}}) \eqqcolon \deg{\mathcal{S}}
\end{equation}
as $\deg{\bar{f}}$ is trivially 1. 

A heuristic explanation for the required condition comes from the observation that after diattenuation, vertically polarized light has a Stokes vector given by
\begin{equation}
    \frac{p_1^2}{1+\cos\kappa} \begin{pmatrix}
    1 & \cos \kappa & 0 & 0 \\ \cos \kappa & 1 & 0 & 0 \\ 0 & 0 & \sin \kappa & 0 \\ 0 & 0 & 0 & \sin \kappa
\end{pmatrix}\begin{pmatrix}
    1 \\ -P \\ 0 \\ 0
\end{pmatrix} \propto \begin{pmatrix}
    1-P\cos\kappa \\ \cos\kappa-P \\0 \\0
\end{pmatrix}.
\end{equation}
Observe that for this to remain vertically polarized, we require $\cos\kappa -P < 0$, or equivalently, $\cos\kappa < P$. As a consequence, if $\tilde{P}(x) > \cos\tilde{\kappa}(x)$ for some $x \in X$, by the smoothness of $\tilde{P}$ and $\tilde{\kappa}$, we can find some open set $V\subseteq X$ such that $\tilde{P}(x) > \cos\tilde{\kappa}(x)$ for all $x \in V$. We can then design an input Skyrmion field where vertically polarized light appears only in that region $\psi^{-1}(V\cap \psi(U))$. The output Skyrmion field will then be without vertically polarized light and, as a non-surjective map into the Poincar\'{e} sphere, is trivially degree 0.

\begin{figure}[!h]
    \centering
    \includegraphics[width=0.97\textwidth]{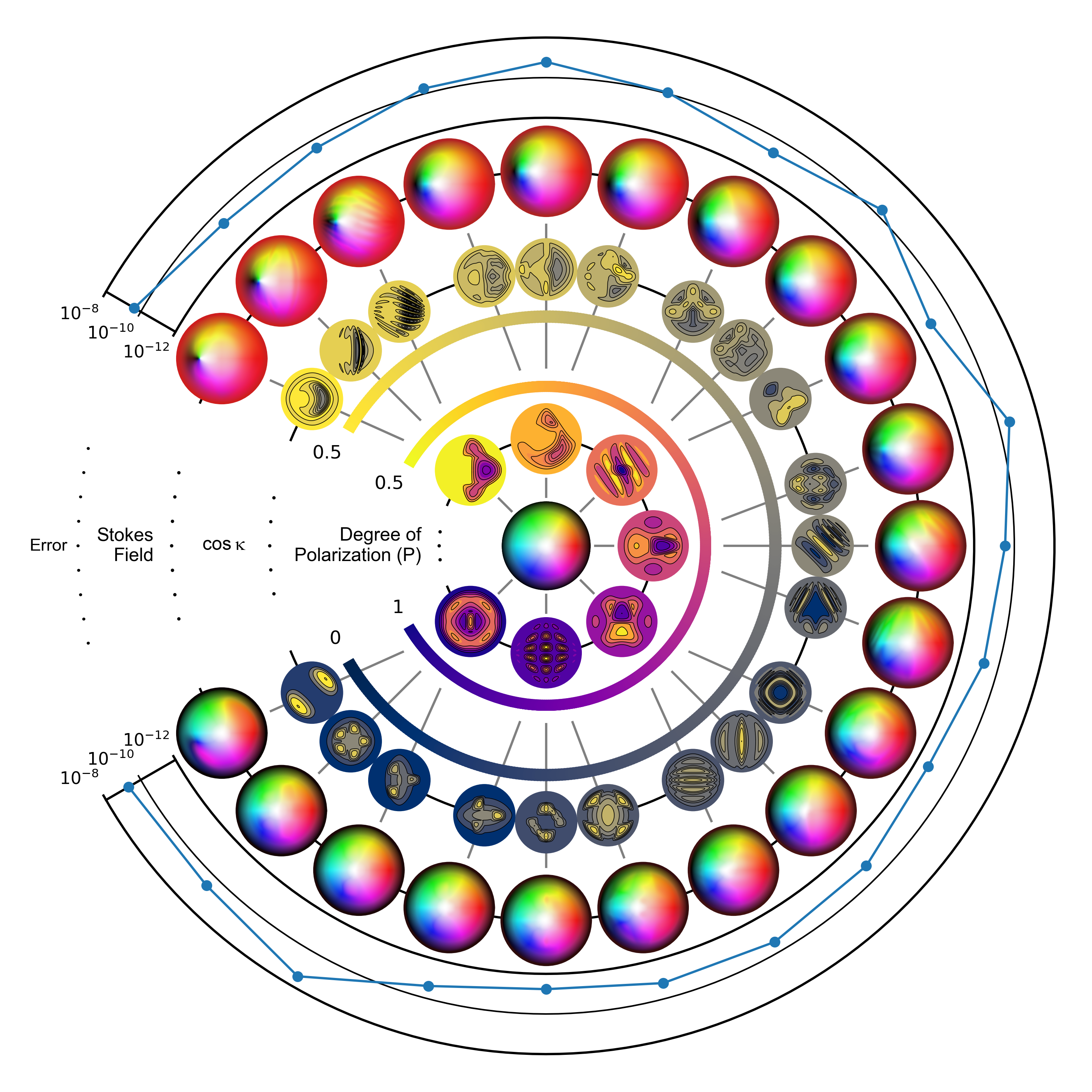}
    \caption{{\bf Skyrmions through spatially varying diattenuators.} The output Stokes field of a first-order N\'{e}el-type Skyrmion with spatially varying degree of polarization after passing through a horizontal linear diattenuator with spatially varying diattenuation. This figure follows a similar structure as Extended Fig. \ref{fig:mixed_retarder}, where the spatial varying parameters $P$ and $\kappa$ are represented by contour plots in the two innermost rings. The Skyrmion numbers are computed using a central difference for partial derivatives and the `nquad' function from SciPy for integration.}
    \label{fig:mixed_diattenuator}
\end{figure}

Extended Fig. \ref{fig:mixed_diattenuator} shows the numerical error in computations for different choices of spatially varying degree of polarization and diattenuation. As in the case of spatially varying retarders, the output Skyrmion number is close to the input value of 1, with errors lying in the range $10^{-8}$ to $10^{-12}$.

\newpage

\section{Non-depolarizing Systems}
In the context of general spatially varying non-depolarizing systems, the singular value decomposition allows us to view the system Jones matrix at every point as a horizontal linear diattenuator sandwiched between two retarders. If, furthermore, this decomposition can be done in a smooth manner, we essentially arrive at a cascade of the two cases presented above and can directly apply the proven conditions. As an application of this, consider the case of a homogeneous non-depolarizing system where such a smooth decomposition is trivially available. Since the constant map extends to a smooth map on $X$, and retarders do not alter the degree of polarization of light, our previous results completely characterize the effects of a homogenous non-depolarizing system on the Skyrmion number of an incoming field in terms of the incident degree of polarization $P$ and the diattenuation of the system $\cos\kappa$, namely 
\begin{equation}
    \deg \mathcal{S}' = \left\{\begin{array}{r l}
        0 & \text{if $\tilde{P}(x) < \cos \kappa$ for all $x\in X$} \\
        \deg \mathcal{S} & \text{if $\tilde{P}(x) > \cos\kappa$ for all $x\in X$}
    \end{array} \right..
\end{equation}

In the extreme case of fully polarized light, the above condition can be elegantly restated in terms of the system's 2\textsuperscript{nd} principal amplitude coefficient,
\begin{equation}
    \deg \mathcal{S}' = \deg \mathcal{S} \quad\quad \text{if $p_2 > 0$}.
\end{equation}
We, therefore, arrive at the powerful result that for fully polarized light passing through a homogeneous non-depolarizing system, the Skyrmion number remains protected unless the system's extinction ratio is simultaneously infinite. 

In practice, things are not so simple, especially when the system's extinction ratio is high. More specifically, the effects of a horizontal linear diattenuator on the Poincar\'{e} sphere become more and more degenerate as diattenuation increases, where a large proportion of the 2-sphere gets mapped onto a small region about $(1,0,0)$ while a small region about $(-1, 0, 0)$ is stretched to cover almost the entire sphere. Qualitatively, this results from the fact that when one eigenstate is attenuated significantly less than the other, its polarization state will dominate the output unless its intensity is also substantially lower. From an experimental perspective, this degeneracy leads to two fundamental difficulties. The first difficulty relates to the limitations of polarimetry at low intensities. Since at high diattenuation, the spatial variation of the output field's polarization state is concentrated in the region of greatest intensity loss, achieving the necessary signal-to-noise ratio to accurately capture this spatial variation is a significant challenge. Secondly, even with perfect measurements, the finite spatial resolution of cameras will still give rise to unavoidable errors when computing the discrete derivatives and numerical integration needed for the Skyrmion number.

These considerations are, of course, not limited to the fully polarized case. In fact, one can visually observe this degeneracy in the upper left quadrant of Extended Fig. \ref{fig:mixed_diattenuator}, where the vertically polarized states become more and more condensed as the incident degree of polarization approaches diattenuation. For a more detailed discussion of these matters, refer to Supplementary Note 4. 

The challenge of experimentally determining the Skyrmion number through spatial sampling highlights the importance of developing alternative ways of identifying the Skyrmion number from polarimetric measurements. From a communications perspective, fast decoding of the Skyrmion number is also essential in maximizing bit-rate, and performing numerical differentiation and integration at the receiver is unlikely to be optimal. One possible property that can be exploited to achieve fast decoding is the fact that the preimage of any regular value under a degree $n$ map is at least of size $\lvert n \rvert$. Using this, we could potentially reduce the evaluation of the complex Skyrmion number integral to a simple bright spot counting problem. A complication with this approach, however, is the fact that there are cases in which it fails --- consider, for instance, a degree $n$ Skyrmion that wraps around the Poincar\'{e} sphere $n+1$ times in one direction and once in the other. Nevertheless, if we transmit an initial Skyrmion that wraps around the Poincar\'{e} sphere exactly $n$ times, such as a standard N\'{e}el or Bloch-type Skyrmion, the field would have to undergo significant deformations before transitioning into one that would result in errors for every polarization state. Therefore, with an appropriate selection of input and channel, it may be feasible to adopt such an intensity-based strategy. 

\section{Depolarizing Media}

In the context of depolarizing media, the problem of topological protection is complicated by the effects of scattering on the propagation of the output Skyrmion field. Nevertheless, when scattering is small but non-zero, a Mueller matrix treatment of the problem is still largely valid. Borrowing the symmetric matrix decomposition proposed by Ossikovski \citesupp{Ossikovski09}, we note that every Mueller matrix $M$ belongs to one of the following two categories:

\begin{enumerate}
    \item Type-I, which can be written $M = M_{J2}M_{\Delta d}M_{J1}$ where $M_{J1}$ and $M_{J2}$ are pure Mueller matrices and 
    \begin{equation}
        M_{\Delta d} = \begin{pmatrix}
            d_0 & 0 & 0 & 0 \\ 0 & d_1 & 0 & 0 \\ 0 & 0 & d_2 & 0 \\ 0 & 0 & 0 & \varepsilon d_3
        \end{pmatrix}
    \end{equation}
    where $0 \leq d_1, d_2, d_3 \leq d_0$ and $\varepsilon \coloneqq \det M/ \lvert \det M\rvert$.

    \item Type-II, which can be written $M = M_{J2}M_{\Delta nd}M_{J1}$ where $M_{J1}$ and $M_{J2}$ are nonsingular pure Mueller matrices and 
    \begin{equation}
        M_{\Delta nd} = \begin{pmatrix}
            2a_0 & -a_0 & 0 & 0 \\ a_0 & 0 & 0 & 0 \\ 0 & 0 & a_2 & 0 \\ 0 & 0 & 0 & a_2
        \end{pmatrix}
    \end{equation}
    where $0\leq a_2 \leq a_0$.
\end{enumerate}

As in the non-depolarizing case, we consider the situation where the spatially varying Mueller matrix can be smoothly decomposed into the above forms. The pure Mueller matrices can then be handled by our previously proven results, so we need only concentrate on $M_{\Delta d}$ and $M_{\Delta nd}$. 

In the type-I case, we show that topological protection is achieved provided $d_1, d_2, d_3 \colon U \longrightarrow \mathbb{R}_{\geq 0}$ are smooth, compactifiable in a compatible way with the incident field (duality condition), satisfy $\text{min}_{x\in X} \{\tilde{d}_1(x), \tilde{d}_2(x), \tilde{d}_3(x)\}>0$ and $\varepsilon$ is everywhere constant. To prove this, we first define the smooth map $f\colon \mathbb{R}^3-\{(0,0,0)\} \longrightarrow S^2$ by 
\begin{equation}
    f(x_1, x_2, x_3) = \frac{1}{\sqrt{x_1^2+x_2^2+x_3^2}}\begin{pmatrix}
        x_1\\ x_2\\ x_3
    \end{pmatrix}
\end{equation}
so that $\mathcal{S}'$ can be written smoothly in terms of $d_1, d_2, d_3$ and $\mathcal{S}$ by 
\begin{equation}
\label{eq:type1depolariser}
    \mathcal{S}'(p) = f(d_1(p)\mathcal{S}(p), d_2(p)\mathcal{S}(p), \varepsilon d_3(p)\mathcal{S}(p))
\end{equation}
We now define a continuous homotopy $H\colon X \times[0,1] \longrightarrow S^2$ as follows
\begin{equation}
    H(x, t) = f\left(\left((1-t)\tilde{d}_1(x)+t\right)\tilde{\mathcal{S}}_1(x),\left((1-t)\tilde{d}_2(x)+t\right)\tilde{\mathcal{S}}_2(x),\left((1-t)\tilde{d}_3(x)+t\right)\varepsilon\tilde{\mathcal{S}}_3(x)\right).
\end{equation}
Notice that since $\tilde{d}_1(x)$, $\tilde{d}_2(x)$, $\tilde{d}_3(x) > 0$ for all $x\in X$, $(1-t)\tilde{d}_i(x)+t > 0$ for all $x\in X$, $t\in [0,1]$ and $i=1,2,3$. Therefore, $H(x,t)$ is well-defined. Moreover, $H(x,0) = \tilde{\mathcal{S}}'(x)$ and $H(x,1) = \tilde{\mathcal{S}}(x)$. Hence, by the usual Whitney Approximation Theorem argument, $\tilde{\mathcal{S}}'$ and $\tilde{\mathcal{S}}$ are smoothly homotopic and 
\begin{equation}
    \deg \mathcal{S}' \coloneqq \deg \tilde{\mathcal{S}}' = \deg \tilde{\mathcal{S}} \eqqcolon \deg \mathcal{S}.
\end{equation}

\begin{figure}[!h]
    \centering \includegraphics[width=0.65\textwidth]{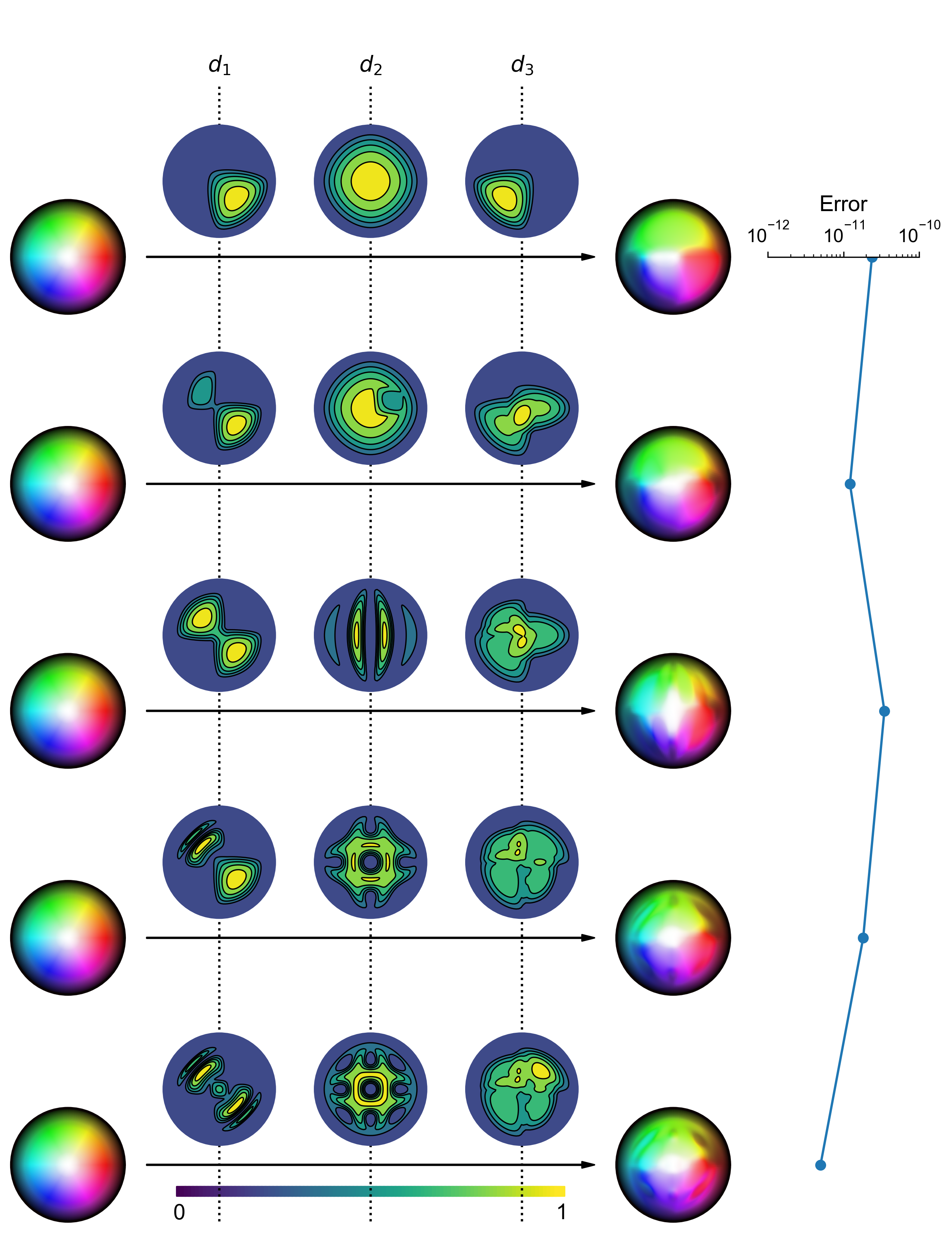}
    \caption{{\bf Skyrmions through spatially varying depolarizers.} The output Stokes field of a first-order N\'{e}el-type Skyrmion after passing through a type-I depolarizing media with spatially varying parameters $d_1, d_2, d_3$ and $\varepsilon = 1$. Note from equation (\ref{eq:type1depolariser}) that the output Stokes field is independent of the incident degree of polarization. This figure therefore does not include the distribution of $P$. The Skyrmion numbers are computed using a central difference for partial derivatives and the `nquad' function from SciPy for integration.}
    \label{fig:depolarising}
\end{figure}

Extended Fig. \ref{fig:depolarising} illustrates the effects of different spatially varying type-I depolarizers on a first-order N\'{e}el-type Skyrmion and the associated numerical error in simulation.

In the type-II case, the induced transformation of $\*M_{\Delta nd}$ on the Poincar\'{e} sphere is given by 
\begin{equation}
    \begin{pmatrix}
        s_1 \\ s_2 \\ s_3
    \end{pmatrix} \mapsto \frac{1}{\sqrt{a_0^2+a_2^2P^2(s_2^2+s_3^2)}}\begin{pmatrix}
        a_0 \\ a_2Ps_2 \\ a_2Ps_3
    \end{pmatrix}
\end{equation}
where $P$ is the incident degree of polarization. Since $0\leq a_0$, any Stokes vector with a negative $s_1$ component is never in the image of this map. Therefore, the output Stokes field of a spatially varying type-II depolarizing media is never surjective, and hence of degree zero. 

\section{Defining the Fractional Skyrmion}
Let $U$ be an open subset of $\mathbb{R}^2$, $X$ a compact, connected, orientable, smooth $2$-dimensional manifold and $\psi \colon U \longrightarrow X$ an orientation preserving diffeomorphism onto its image. Then for any smooth map $\alpha\colon U \longrightarrow Y$, we say $\alpha$ is $\psi$-extendable if 

\begin{enumerate}
    \item $\alpha \circ \psi^{-1}$ extends via continuity to $\overline{\psi(U)}$ and
    \item there exists a smooth $F\colon X-\psi(U) \longrightarrow Y$ such that $F\lvert_{\partial \psi(U)} = \alpha\circ\psi^{-1}\lvert_{\partial \psi(U)}$, where the latter is to be understood as the continuous extension of $\alpha \circ \psi^{-1}$ to $\overline{\psi(U)}$.
\end{enumerate}
We then call
\begin{equation}
    \tilde{\alpha}_F = \left\{\begin{array}{r l}
        \alpha \circ \psi^{-1}, & x\in\overline{\psi(U)} \\
        F, & x\in X-\psi(U) 
    \end{array}\right.
\end{equation}
the $(\psi, F)$-extension of $\alpha$ to $X$, where $\tilde{\alpha}_F$ is continuous by a gluing argument and piecewise smooth by construction. 

Now suppose $\mathcal{S}\colon U \longrightarrow S^2$ is $\psi$-extendable and let $\tilde{\mathcal{S}}_F$ be a $(\psi,F)$-extension. Then $\tilde{\mathcal{S}}_F$, as a continuous and piecewise smooth function between compact, connected, orientable $2$-dimensional manifolds, can be assigned a degree
\begin{equation}
\label{eq: degree and fractional relationship}
    \deg \tilde{\mathcal{S}}_F = \int_X \tilde{\mathcal{S}}_F^\ast\omega_0 = \text{SkyN}(\mathcal{S}) + \int_{X-\psi(U)} F^\ast\omega_0.
\end{equation}
Notice that this is a strict generalization of our previous notion of the Skyrmion, where $F$ is uniquely defined by continuity whenever $\psi(U)$ is dense so that $\tilde{\mathcal{S}}$ agrees with its only possible $(\psi,F)$-extension $\tilde{\mathcal{S}}_F$.

\section{Interpreting Fractional Skyrmions}

\begin{figure}[ht]
    \centering
    \includegraphics[width=0.95\textwidth]{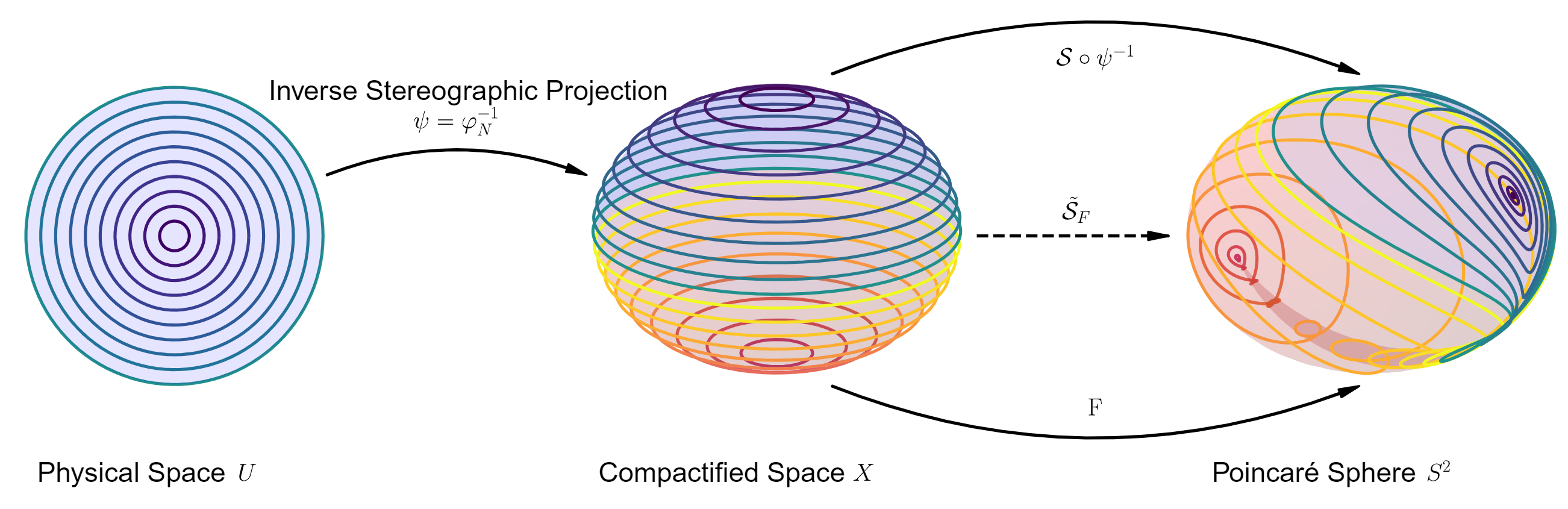}
    \caption{{\bf Geometric interpretation of fractional Skyrmions.} (Left) Physical space $U$ with reference lines. (Middle) Compactified space $X$ with showing reference lines after stereographic projection and fictitious lines corresponding to $X-\psi(U)$. (Right) Poincar\'{e} sphere with the image of the reference lines under $\tilde{\mathcal{S}}_F$}.
    \label{fig: geometric fractional Skyrmions}
\end{figure}

In this section, we discuss two ways in which one can understand the definition of fractional Skyrmions given in this paper. To begin, as an example of the extension procedure presented, consider the situation $U = B_R(0)$. In this case, we may take $\psi(x,y) = \varphi_N^{-1}(x/R, y/R)$ where $\varphi_N$ is the stereographic projection map from the south pole so that $U$ is mapped to the upper hemisphere $S^+ = \{(x^1, x^2, x^3)\in S^2 \colon x^3>0\} \subseteq X = S^2$ as shown in Extended Fig. \ref{fig: geometric fractional Skyrmions}. Now suppose $\mathcal{S}\circ \psi^{-1}$ extends via continuity to $\overline{S^+}$ and let $\partial \tilde{\mathcal{S}}$ be this extension when restricted to equator $\partial S^+$. Then $\partial\tilde{\mathcal{S}}\colon \partial S^+ \cong S^1 \longrightarrow S^2$, as a non-surjective map into $S^2$, is null-homotopic. For example, if $-q \in S^2$ is not in the image of $\partial\tilde{\mathcal{S}}$, then $H\colon \partial S^+ \times [0,1] \longrightarrow S^2$ given by 
\begin{equation}
\label{eq: homotopy of circle}
    H(\theta,t) = \frac{(1-t)\partial\tilde{\mathcal{S}}(\theta)+tq}{\lVert (1-t)\partial\tilde{\mathcal{S}}(\theta)+tq \rVert}
\end{equation}
is a possible null-homotopy between $\partial\tilde{\mathcal{S}}$ and the constant map taking value $q$. In general, let $H\colon \partial S^+ \times [0,1] \longrightarrow S^2$ be a homotopy from $\partial\tilde{\mathcal{S}}$ to some $q\in S^2$. We may then use this null-homotopy to extend $\mathcal{S}\circ \psi^{-1}$ to the lower hemisphere $S^- = \{(x_1, x_2, x_3)\in S^2\colon x_3\leq 0\}$ by $F\colon S^- \longrightarrow S^2$,
\begin{equation}
\label{eq: nullhomotopy and extension}
    F\left(\sqrt{1-x_3^2}\cos\theta, \sqrt{1-x_3^2}\sin\theta, x_3\right) = H(\theta, -x_3),
\end{equation}
noting that this is well-defined as $H$ descends onto the quotient $\partial S^+\times [0,1]/{\sim}$ given by the equivalence relation $(\theta_1, 1) \sim (\theta_2, 1)$ for all $\theta_1,\theta_2 \in \partial S^+$, and this space is clearly homeomorphic to $S^-$ via $[\theta, z] \mapsto (\sqrt{1-z^2}\cos\theta, \sqrt{1-z^2}\sin\theta, -z)$. Conversely, given any well-defined $F$, equation (\ref{eq: nullhomotopy and extension}) defines a null-homotopy of $\partial \tilde{\mathcal{S}}$. Therefore, in this context, we may think of a $(\psi, F)$-extension as a specific choice of null-homotopy of $\partial \tilde{S}$, and the topological character of such fractional Skyrmions is expressed in the restrictions to the functions $F$ that extend them. Extended Fig. \ref{fig: geometric fractional Skyrmions} shows an example of $F$ constructed with the homotopy defined by equation (\ref{eq: homotopy of circle}). Note also that a similar approach can be used to extend $\mathcal{S}$ for any connected $U$, where appropriate null-homotopies can be used to fill in missing holes on $X=S^2$.  

\begin{figure}[ht]
    \centering
    \includegraphics[width=0.85\textwidth]{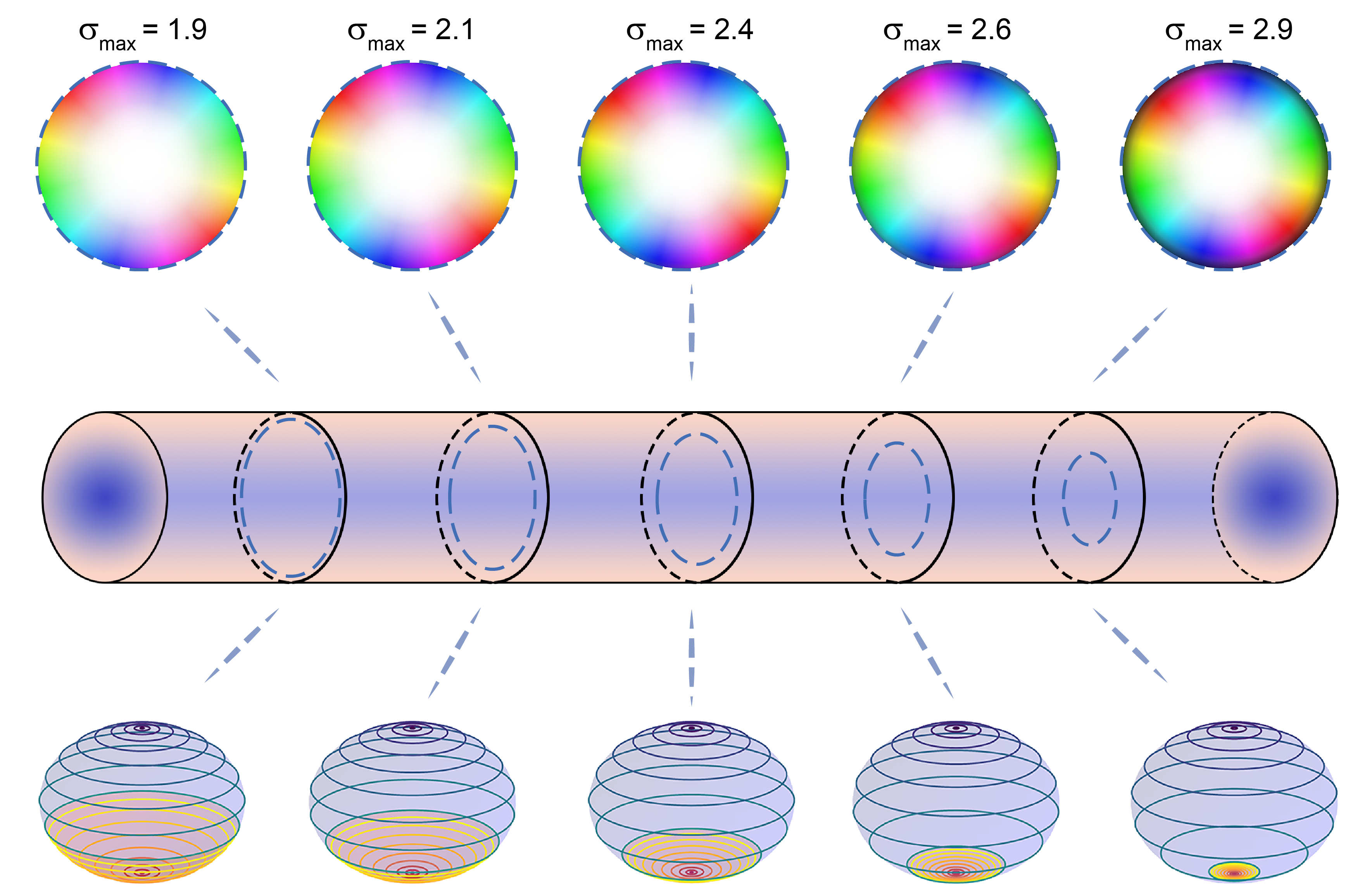}
    \caption{{\bf Fractional Skyrmions in a gradient index lens.} The propagation of right circularly polarized light through a gradient index lens system and the respective extensions of the Stokes field given by equation (\ref{eq: homotopy of circle}) taking $q$ to be $(0,0,-1)$. Note we assume that all fields lie within a single pitch of the system.} 
    \label{fig: fractional Skyrmions through GRIN lens}
\end{figure}

Another way of understanding equation (\ref{eq: degree and fractional relationship}) is that the function $F$ exists to provide a correction term $\int_{X-\psi(U)} F^\ast \omega_0$ which accounts for perturbations of the boundary. This is most evident when considering the propagation of right circularly polarized light through a gradient index lens system as shown in Extended Fig. \ref{fig: fractional Skyrmions through GRIN lens}. A most remarkable property of the gradient index lens is that when uniform right circularly polarized light $s = (1,0,0)^T$ couples into a gradient index lens with $\sigma_{\max} = \pi$, the output is a N\'{e}el-type Skyrmion of degree 2 (See Supplementary Note 5 for the mathematical details). However, for any $\sigma_{\max} < \pi$, the output field is not compactifiable. Nonetheless, when $\sigma_{\max}$ is close to $\pi$, the output field closely resembles that of a N\'{e}el-type Skyrmion, and intuitively, its Skyrmion density integrates to a value close to $2$. The extension procedure, as shown in Extended Fig. \ref{fig: fractional Skyrmions through GRIN lens}, can then be seen as a way of canonically assigning values so that the resultant map from $X$ is simply that of a N\'{e}el type Skyrmion. In this way, one can view such ``fractional Skyrmions'' as a fraction of a regular Skyrmion obtained through extension, hence motivating its name.

\section{The Topological Protection of Fractional Skyrmions}

Through equation (\ref{eq: degree and fractional relationship}), we can relate the Skyrmion number of a fractional Skyrmion with the degree of a $(\psi, F)$-extension of $\mathcal{S}$. In this way, we can make use of the homotopy invariance of the degree to establish robustness results similar to those of regular Skyrmions. We proceed as follows. Let $\mathcal{S}' \colon U \longrightarrow S^2$ be the output Stokes field and $\tilde{\mathcal{S}}'_{F'}$ a $(\psi, F')$-extension of $\mathcal{S'}$. If we can construct a homotopy from $\tilde{\mathcal{S}}'_{F'}$ to a map of the form $\bar{f} \circ \tilde{\mathcal{S}}_F$ for some smooth $\bar{f}$ then
\begin{align}
    \text{SkyN}(\mathcal{S}')+\int_{X-\psi(U)}F'^\ast\omega_0 = \deg \tilde{\mathcal{S}}'_{F'} & = \deg(\bar{f}\circ\tilde{\mathcal{S}}_F) \nonumber \\ &= \deg(\bar{f})\deg(\tilde{\mathcal{S}}_F) = \deg(\bar{f})\left(\text{SkyN}(\mathcal{S})+\int_{X-\psi(U)}F^\ast\omega_0\right)
\end{align}
so that 
\begin{equation}
\label{eq: Fractional Skyrmion Protection}
    \text{SkyN}({\mathcal{S}'}) = \deg(\bar{f})\text{SkyN}({\mathcal{S}}) + \int_{X-\psi(U)} (\deg(\bar{f})F-F')^\ast\omega_0.
\end{equation}

An important situation where equation (\ref{eq: Fractional Skyrmion Protection}) simplifies is when $(\mathcal{S}'\circ \psi^{-1})(x) = (f\circ\mathcal{S}\circ\psi^{-1})(x)$ for every $x \in \partial \psi(U)$ where $f(s) = Rs$ for some $R\in SO(3)$. Note here that we understand $\mathcal{S}\circ\psi^{-1}$ and $\mathcal{S}'\circ\psi^{-1}$ in the previous equation as continuous extensions onto $\overline{\psi(U)}$, and whose existence is guaranteed by the conditions in our definition of $\psi$-extendability. One can then readily verify that $f^\ast \omega_0 = \omega_0$, so if we take $F' = f\circ F$, 
\begin{equation}
    \int_{X-\psi(U)} F'^\ast \omega_0 = \int_{X-\psi(U)} F^\ast(f^\ast \omega_0) = \int_{X-\psi(U)}F^\ast\omega_0,
\end{equation}
which implies
\begin{equation}
    \text{SkyN}(\mathcal{S}') = \deg(\bar{f})\text{SkyN}(\mathcal{S}) + \left(\deg(\bar{f})-1\right)\int_{X-\psi(U)} F^\ast\omega_0.
\end{equation}

With the simplification above, we may arrive at similar topological protection results for fractional Skyrmions as stated below. 
\begin{enumerate}
    \item A spatially varying retarder $e^{i\phi(p)}J(p)$ preserves the Skyrmion number whenever $J$ is smooth and $(J\circ \psi^{-1})(x)$ is constant for all $x\in \partial\psi(U)$.
    \item The Skyrmion number of a field with spatially varying degree of polarization $P$ impinging on a spatially varying diattenuator with diattenuation $\cos \kappa$ remains unchanged if $P$, $\cos \kappa$ are smooth and $P\circ\psi^{-1}\lvert_{\partial\psi(U)} = 1$, $\kappa\circ\psi^{-1}\lvert_{\partial\psi(U)} = \pi/2$.
    \item The Skyrmion number of a field with spatially varying degree of polarization $P$ impinging on a spatially varying type-I depolarizer remains unchanged if $P$, $d_1$, $d_2$, $d_3$ are smooth, $\varepsilon=1$ and $d_1\circ\psi^{-1}\lvert_{\partial \psi(U)}=d_2\circ\psi^{-1}\lvert_{\partial \psi(U)}=d_3\circ\psi^{-1}\lvert_{\partial \psi(U)}$.  
\end{enumerate}

Notice that these are somewhat weaker conditions than those established for ordinary Skyrmions, and this is what we expect. By broadening the notion of the Skyrmion, a larger class of functions is considered, and hence, the transformations that leave the Skyrmion number of every field unchanged necessarily become more restrictive.

\clearpage
\bibliographystylesupp{unsrt}
\bibliographysupp{main}

\end{document}


\maketitle
\section{The Integral Formula for Skyrmion Number}
To evaluate the pullback 
\begin{equation}
    \mathcal{S}^\ast\omega_0 = \frac{1}{4\pi}(\iota \circ S)^\ast (x^1 dx^2\wedge dx^3 - x^2 dx^1\wedge dx^3 + x^3 dx^1 \wedge dx^2),
\end{equation}
let $\iota \circ \mathcal{S} = S = (S^1, S^2, S^3)$ so that $x^i \circ S = S^i$, $i=1,2,3$. Then, by direction computation, 
\begin{align}
    (\iota \circ \mathcal{S})^\ast (x^1 dx^2\wedge dx^3) & = S^1\left(\frac{\partial S^2}{\partial x}dx + \frac{\partial S^2}{\partial y}dy\right)\wedge\left(\frac{\partial S^3}{\partial x}dx+\frac{\partial S^3}{\partial y}dy\right) \nonumber \\
    & = S^1\left(\frac{\partial S^2}{\partial x}\frac{\partial S^3}{\partial y} - \frac{\partial S^2}{\partial y}\frac{\partial S^3}{\partial x}\right) dx\wedge dy. 
\end{align}
Likewise, 
\begin{align}
    (\iota \circ S)^\ast(-x^2 dx^1 \wedge dx^3) & = -S^2\left(\frac{\partial S^1}{\partial x}dx + \frac{\partial S^1}{\partial y}dy\right)\wedge\left(\frac{\partial S^3}{\partial x}dx+\frac{\partial S^3}{\partial y}dy\right) \nonumber \\
    & = S^2\left(\frac{\partial S^3}{\partial x}\frac{\partial S^1}{\partial y} - \frac{\partial S^3}{\partial y}\frac{\partial S^1}{\partial x}\right) dx\wedge dy. 
\end{align}
and 
\begin{align}
    (\iota \circ S)^\ast (x^3 dx^1\wedge dx^2) & = S^3\left(\frac{\partial S^1}{\partial x}dx + \frac{\partial S^1}{\partial y}dy\right)\wedge\left(\frac{\partial S^2}{\partial x}dx+\frac{\partial S^2}{\partial y}dy\right) \nonumber \\
    & = S^3\left(\frac{\partial S^1}{\partial x}\frac{\partial S^2}{\partial y} - \frac{\partial S^1}{\partial y}\frac{\partial S^2}{\partial x}\right) dx\wedge dy. 
\end{align}
Therefore, 
\begingroup
\small
\begin{align}
    \deg(\mathcal{S}) & = \frac{1}{4\pi}\int_U \left[S^1\left(\frac{\partial S^2}{\partial x}\frac{\partial S^3}{\partial y} - \frac{\partial S^2}{\partial y}\frac{\partial S^3}{\partial x}\right) + S^2\left(\frac{\partial S^3}{\partial x}\frac{\partial S^1}{\partial y} - \frac{\partial S^3}{\partial y}\frac{\partial S^1}{\partial x}\right) + S^3\left(\frac{\partial S^1}{\partial x}\frac{\partial S^2}{\partial y} - \frac{\partial S^1}{\partial y}\frac{\partial S^2}{\partial x}\right) \right]dx\wedge dy \nonumber \\ &= \frac{1}{4\pi}\iint_U S \cdot \left(\frac{\partial S}{\partial x} \times \frac{\partial S}{\partial y} \right) dx dy.
\end{align}
\endgroup

\section{The Whitney Approximation Theorem}
The Whitney Approximation Theorem is an important tool that allows us to turn continuous homotopies into smooth homotopies. We give a brief account of the key ideas here. To begin, we make the following definition. Let $M$ and $N$ be smooth manifolds and $A \subseteq M$ a closed subset of $M$. We say $g\colon M \longrightarrow N$ is smooth on $A$ if there exists an open set $U \supseteq A$ and a smooth function $g_0\colon U \longrightarrow N$ such that $g_0\lvert_A = g\lvert_A$. 

The Whitney Approximation Theorem is then the following statement. Let $M$ and $N$ be smooth manifolds, $A$ a closed subset of $M$, and $g\colon M \longrightarrow N$ a continuous map that is smooth on $A$. Then there exists a smooth map $f\colon M \longrightarrow N$ homotopic to $g$ such that $f\lvert_A = g\lvert_A$.  

An immediate corollary to this is that any smooth maps $f_0, f_1 \colon M \longrightarrow N$ that are homotopic are also smoothly homotopic. To see this, let $F\colon M \times[0,1] \longrightarrow N$ be a homotopy connecting $f_0$ and $f_1$. We may continuously extend $F$ to a mapping $\tilde{F}\colon M \times \mathbb{R}$ by $\tilde{F}(x,t) = F(x,0)$ when $t\leq 0$ and $\tilde{F}(x,t) = F(x,1)$ when $t\geq 1$. Then $\tilde{F}$ is a continuous map from $M\times \mathbb{R}$ to $N$. Moreover, $\tilde{F}$ is trivially smooth on $M \times \{0\}$ and $M\times \{1\}$ (We can, for instance, take $\tilde{F}_0\colon M \times (-\epsilon, \epsilon)$, $\tilde{F}_0(x,t) = F(x,0)$, then $\tilde{F}_0$ is clearly smooth and $\tilde{F}_0\lvert_{M\times\{0\}} = \tilde{F}\lvert_{M\times\{0\}}$, and similarly for $M\times\{1\}$). The result then follows by direct application of the Whitney Approximation Theorem.

\section{N\'{e}el and Bloch-type Skyrmions}
Within our framework, we can generalize N\'{e}el and Bloch-type Skyrmions to arbitrary domains $U$ in the following way. We begin by constructing a canonical collection of maps $\tilde{\mathcal{S}}$ of arbitrary degree. Note first that if $g \colon S^1 \longrightarrow S^1$ is a map of degree $n$ and $\Sigma \colon \text{Top} \longrightarrow \text{Top}$ the suspension functor, then a simple argument via the Mayer-Vietoris sequence shows that $\Sigma g \colon \Sigma S^1 \cong S^2 \longrightarrow \Sigma S^1 \cong S^2$ is also of degree $n$. Now, for any $n \in \mathbb{Z}$ and $\gamma \in \mathbb{R}$, let $g_{n, \gamma}\colon S^1 \longrightarrow S^1$ be $g_{n, \gamma}(z)=e^{{i\mkern1mu}\gamma}z^n$. As $\deg g_{n, \gamma}$ is easily checked to be $n$,
\begin{equation}
    \Sigma g_{n, \gamma}(\sin\theta\cos\varphi, \sin\theta\sin\varphi, \cos\theta) = (\sin\theta\cos (n\varphi+\gamma), \sin\theta \sin (n\varphi+\gamma), \cos\theta)
\end{equation}
is a map from $S^2$ to $S^2$ of degree $n$. We can then use these to define a family of optical fields $S_{n, \gamma}\colon U \longrightarrow S^2$ given by $S_{n, \gamma} = \Sigma g_{n, \gamma} \circ \psi$ such that $\deg S_{n,\gamma} = n$. In this construction, $\gamma = 0$ corresponds to the usual N\'{e}el-type Skyrmions while $\gamma = \pi/2$ gives the usual Bloch-type Skyrmions. 

\section{The Observability of Skyrmion Fields at High Diattenuation}

\begin{figure}[ht]
\begin{center}
\includegraphics[width = 0.96\textwidth]{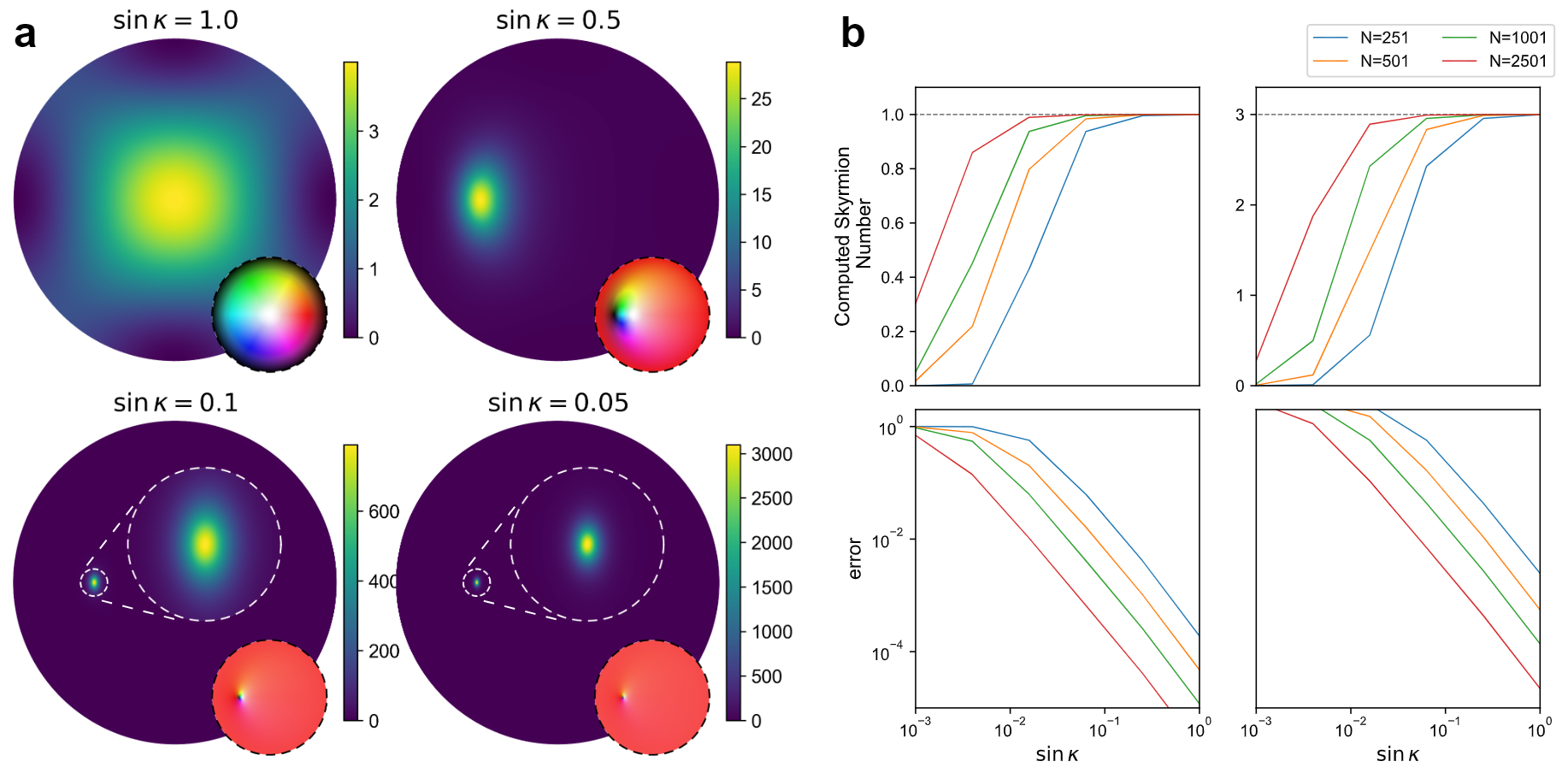}
\end{center}
\caption{{\bf Simulated Skyrmion number integral at increasing levels of diattenuation.} {\bf a}, The Skyrmion density and Stokes field of a fully polarized first-order N\'{e}el-type Skyrmion after
passing through horizontal linear diattenuators of different diattenuation. {\bf b}, Numerically evaluated Skyrmion number and the associated error of a (left) first-order and (right) third-order N\'{e}el-type Skyrmion after propagation through horizontal linear diattenuators of different diattenuation.}
\label{fig: supplement 2}
\end{figure}

As mentioned in the main text, the degeneracy of the induced mapping on the Poincar\'{e} sphere at high diattenuation causes challenges in the numerical evaluation of the Skyrmion number. Graphically, this can be understood from the observation that the Skyrmion density approaches a delta function as diattenuation increases (Supplementary Fig. \ref{fig: supplement 2}a), and this imposes additional requirements on spatial resolution akin to the usual Nyquist-like considerations for accurate computation of the Skyrmion number. In a real polarimetry set-up, the finite spatial resolution of cameras means that we only have access to a spatially averaged Stokes vector at specified points, and this limits the ways in which we can compute numerical derivatives and numerical integrals. We can, however, emulate these numerical errors by sampling a theoretical Skyrmion field $N\times N$ times and computing the Skyrmion number integral from these samples. In the following, we use a central difference to estimate the derivative and Simpson's rule for integration. In all simulations, we consider fully polarized incident light passing through a horizontal linear diattenuator. 

The results of our simulations are shown in Supplementary Fig. \ref{fig: supplement 2}b. As predicted, the grid size needed to obtain a value close to the true Skyrmion number increases with diattenuation. Additionally, for the same grid size, the more complicated third-order Skyrmion fields tend to have poorer accuracy in comparison to their first-order counterparts, reflecting a potential challenge in the detection of high-order Skyrmions. Nonetheless, the results are promising. From a practical perspective, polarisers typically have an extinction ratio in the range $100 \colon 1$ to $10^6 \colon 1$, which corresponds to $0.002 < \sin \kappa < 0.198$. While the lower end of this range requires a grid size on the order $N \gtrsim 1000$ for accurate computations, which may well be a challenge to realize in practice, the higher end can be handled by standard-resolution cameras. We stress that these results only account for the accuracy of numerical calculations and not scattering, noise, and other potential experimental imperfections.

\section{Skyrmion Number Integral in Gradient Index Systems}

In this section, we use gradient index lens systems to demonstrate the importance of compactifiability in our definition of the Skyrmion. Consider a gradient index lens, whose effect on polarisation is described by a contraction followed by composition with the smoothly varying but non-compactifiable Mueller matrix 
\begin{equation}
    M(r, \theta) = \begin{pmatrix}
        1 & 0 & 0 & 0 \\ 0 & \cos^22\theta+\sin^22\theta\cos\sigma(r) & \sin 2\theta \cos 2\theta (1-\cos \sigma(r)) & -\sin 2\theta\sin \sigma(r)  \\ 0 & \sin 2\theta \cos 2\theta (1-\cos \sigma(r)) & \sin^2 2\theta + \cos^2 2\theta \cos\sigma(r) & \cos 2\theta \sin \sigma(r) \\ 0 & \sin 2\theta \sin \sigma(r) & -\cos2\theta \sin\sigma(r) & \cos \sigma(r)
    \end{pmatrix},
\end{equation}
where $\sigma(r)$ smoothly increases from 0 at $r=0$ to some $\sigma_{\max}$ at the outer circumference $r=r_0$. In the following derivation, we show that for a homogeneous input $s=(s_1, s_2, s_3)$, the Skyrmion number integral evaluates to 
\begin{equation}
    (1-\cos \sigma_{\max}) s_3,
\end{equation}
which is not necessarily an integer. Notice further that when $\sigma_{\max} = 2n\pi$, $M$ is constant on the boundary. In this case, the output Skyrmion number evaluates to 0 for all homogeneous inputs, in agreement with the topological protection results established in this paper. 

Our main approach to simplifying the Skyrmion number integral is through the use of Stokes' theorem. First, note that with respect to the usual spherical coordinates $(\sin\theta\cos\phi, \sin\theta\sin\phi, \cos\theta)$, we may write
\begin{equation}
    \omega_0 = \sin\theta d\theta \wedge d\phi.
\end{equation}
Letting $N=(0,0,1)$ be the north pole and $S=(0,0,-1)$ be the south pole, notice that on $S^2 - \{N,S\}$, the 1-form $\eta = -\cos\theta d\phi$ is easily seen to satisfy $d\eta = \omega_0$. Therefore, for any $\mathcal{S}\colon U \longrightarrow S^2 - \{N,S\}$, 
\begin{equation}
    d(\mathcal{S}^\ast\eta) = S^\ast(d\eta) = \mathcal{S}^\ast\omega_0.
\end{equation}

We can, therefore, define a Skyrmion potential by 
\begin{equation}
    \mathcal{A} = \mathcal{S}^\ast\eta = S_3d\left(\tan^{-1}\left(\frac{S_1}{S_2}\right)\right).
\end{equation}
Stokes' theorem then gives 
\begin{equation}
\label{eq: Stokes for Skyrmion number}
    \text{SkyN}(\mathcal{S}) = \frac{1}{4\pi} \int_U \mathcal{S}^\ast \omega_0 = \frac{1}{4\pi} \int_U d(\mathcal{S}^\ast \eta) = \frac{1}{4\pi} \int_{\partial U} \iota^\ast \mathcal{A},
\end{equation}
where $\iota \colon \partial U \hookrightarrow U$ is the inclusion map and $\partial U$ having the induced orientation. Specializing to the gradient index lens, we may take $U = B_{r_0}(0) \subset \mathbb{R}^2\cong \mathbb{C}$ with its standard orientation. Then for any $\mathcal{S}\colon U \longrightarrow S^2 -\{N,S\}$,
\begin{equation}
    \text{SkyN}(\mathcal{S}) = \frac{1}{4\pi}\int_0^{2\pi} S_3(r_0e^{i \theta}) \frac{d}{d\theta} \tan^{-1}\left(\frac{S_1(r_0e^{i\theta})}{S_2(r_0e^{i\theta})} \right) d\theta.
\end{equation}

 In the case where $S_3$ takes values $\pm 1$ at some point $p\in U$. We may still define $\mathcal{A}$ everywhere else and evaluate the Skyrmion number by running a contour around $p$ as in Supplementary Fig. \ref{fig: Contour Example}a. The contribution of the singularity at $p$ can then be computed by 
\begin{align}
    \lim_{\varepsilon \searrow 0} & -\frac{1}{4\pi} \int_{0}^{2\pi} S_3(p+\varepsilon e^{i\theta})\frac{d}{d\theta} \tan^{-1}\left(\frac{S_1(p+\varepsilon e^{i\theta})}{S_2(p+\varepsilon e^{i\theta})} \right) d\theta \nonumber \\ 
    & = -\frac{S_3(p)}{2} \times \lim_{\varepsilon \searrow 0} \left(\text{winding number of $(S_2(p+\varepsilon e^{i\theta}), S_1(p+\varepsilon e^{i\theta}))$ about $(0,0)$}\right).
\end{align} 
With this machinery on hand, we can break the problem down into three separate cases. We focus on the situation where $0<\sigma_{\max} \leq \pi$, noting that for $\sigma_{\max} > \pi$, $U$ can be divided down into the disk given by $0 \leq r \leq \sigma^{-1}(\pi)$ and annuli given by $\sigma^{-1}(\pi) \leq r \leq \sigma^{-1}(2\pi)$, $\ldots$, $\sigma^{-1}(k\pi) \leq r \leq r_0$, which can be individually handled in the manner described below. 
\begin{figure}[ht]
    \centering
    \includegraphics[width=0.95\textwidth]{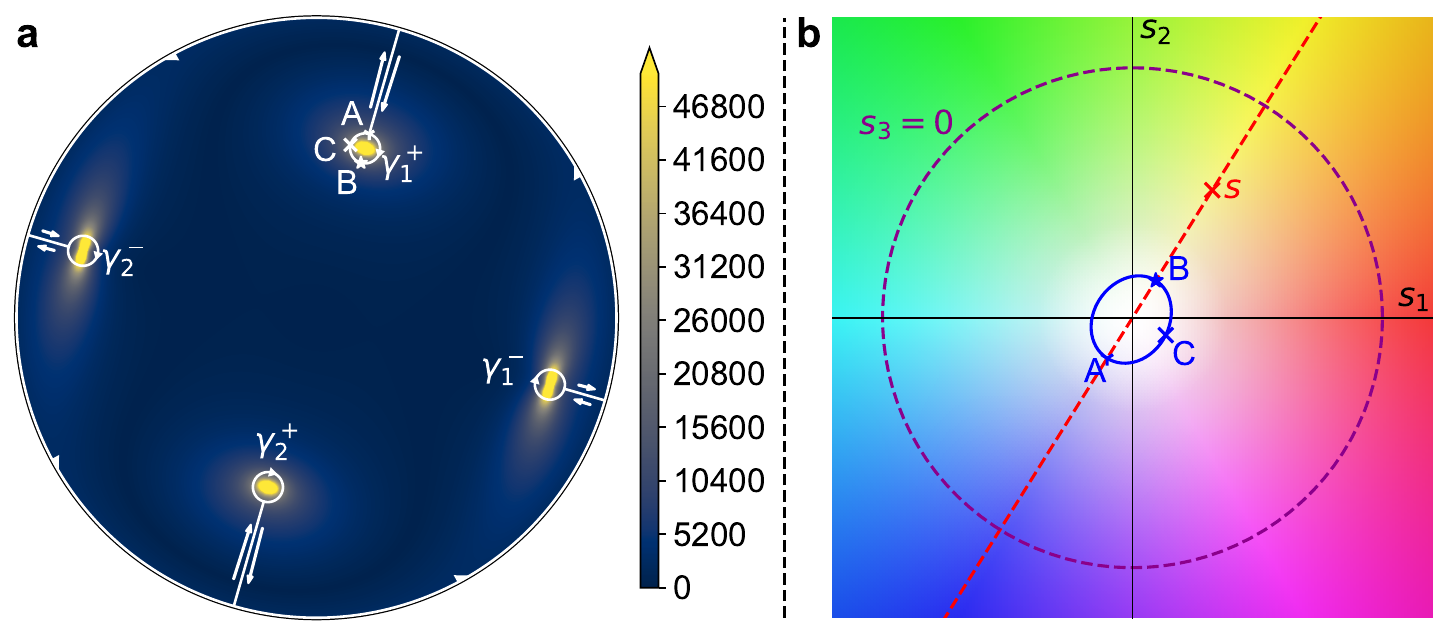}
    \caption{{\bf Contour for Skyrmion number integral.} {\bf a}, Magnitude of $\mathcal{A}(r, \theta)$ for a gradient index lens system with input $(s_1, s_2, s_3) \propto (1, 1, 1)$, $\sigma_{\max} = \pi$ and birefringence quadratic with radius. White lines show a positively oriented contour that can be used to evaluate the Skyrmion number. {\bf b}, The image of $\gamma_1^+$ under $\mathcal{S}$ after stereographic projection from the south pole along with distinguished points used to determine the winding number of $\kappa_1^+$.} 
    \label{fig: Contour Example}
\end{figure}

\subsection{\texorpdfstring{Case I: $\bm{s_3 = \pm 1}$}{TEXT}}

When $s_3 = \pm 1$, the output Stokes field is given by  
\begin{equation}
    \iota \circ \mathcal{S} = \begin{pmatrix}
        S_1 \\ S_2 \\ S_3
    \end{pmatrix} = \begin{pmatrix}
        -s_3\sin\sigma \sin 2\theta \\ s_3\sin \sigma \cos 2\theta \\ s_3\cos \sigma
    \end{pmatrix}
\end{equation}
so that $\mathcal{A} = -2s_3\cos\sigma d\theta$, and where this is well-defined for all $0<r\leq r_0$ when $\sigma_{\max} < \pi$ and $0<r<r_0$ when $\sigma_{\max} = \pi$. We may therefore directly apply equation (\ref{eq: Stokes for Skyrmion number}) to get
\begin{align}
    \text{SkyN}(\mathcal{S}) = \lim_{\varepsilon \searrow 0} \frac{1}{4\pi} \int_0^{2\pi} -2s_3& \left(\cos(\sigma(r_0-\varepsilon)) -\cos(\sigma(\varepsilon))\right) d\theta \nonumber  \\ & \qquad \qquad = -s_3\left(\cos(\sigma(r_0)) - \cos(\sigma(0))\right) = (1-\cos\sigma_{\max})s_3
\end{align}

\subsection{\texorpdfstring{Case II: $\bm{-1 < s_3 < 1,\, \sigma_{\max} = \pi}$}{TEXT}}
Notice the output stokes field when restricted to $r = r_0$ is 
\begin{equation}
    \begin{pmatrix}
        S_1(r_0e^{i\theta}) \\ S_2(r_0e^{i\theta}) \\ S_3(r_0e^{i\theta})
    \end{pmatrix} = \begin{pmatrix}
        s_1\cos 4\theta + s_2\sin 4\theta \\ s_1\sin 4\theta - s_2\cos 4\theta \\ -s_3
    \end{pmatrix}
\end{equation}
Therefore, 
\begin{equation}
    \frac{1}{4\pi} \int_{\partial U} \iota^\ast \mathcal{A} = \frac{-s_3}{2} \times \left(\text{winding number of $( s_1\sin 4\theta -s_2\cos 4\theta, s_1\cos 4\theta + s_2\sin 4\theta)$ about $(0,0)$} \right)
\end{equation}

Setting $\tan \phi = s_1/s_2$, we see that
\begin{align}
    s_1\sin 4\theta - s_2\cos 4\theta & = -\sqrt{s_1^2 + s_2^2} \cos(4\theta + \phi) \\
    s_1\cos 4\theta+s_2\sin 4\theta & = \sqrt{s_1^2 + s_2^2} \sin (4\theta + \phi) 
\end{align}
so the winding number of $(s_1\sin 4\theta -s_2\cos 4\theta, s_1\cos 4\theta + s_2\sin 4\theta)$ about $(0,0)$ is easily seen to be $-4$, independent of $s$. Hence 
\begin{equation}
    \frac{1}{4\pi} \int_{\partial U} \iota^\ast \mathcal{A} = 2s_3.
\end{equation}

We now need to account for the points $S_3 = \pm 1$. We first consider $S_3 = 1$. To begin, notice that $M(r,\theta)$ induces a rotation of angle $\sigma(r)$ about the unit axis $(-\cos 2\theta, -\sin 2\theta, 0)$. On the other hand, the required transformation from $(s_1, s_2, s_3)$ to $(0,0,1)$ is a rotation about the axis $(s_2, -s_1, 0)$ by an angle $\cos \phi = s_3$. By matching these conditions, we may solve for $S_3 = \pm 1$. However, the advantage of adopting the Skyrmion potential is that we deal with winding numbers, so we need not solve these equations explicitly. Indeed, doing so is impossible without an explicit formula for $\sigma(r)$. The strength of dealing with topological concepts is that a qualitative picture is sufficient.

To see how we may proceed, note first that as $\sigma(r)$ is strictly increasing from $0$ to $\pi$, $\cos\sigma(r)$ is strictly decreasing from $1$ to $-1$, and there is therefore a single $\rho_+$ such that $\cos(\sigma(\rho_+)) = s_3$. As for the equation in $\theta$, it is easy to see that there are precisely two solutions, which we denote $\xi$ and $\xi + \pi$. Therefore, we need to determine the winding numbers of the two curves $\kappa_1^+$ and $\kappa_2^+$ defined by 
\begin{align}
    \kappa_1^+(\chi) & = (S_2(\gamma_1^+(\chi)), S_1(\gamma_1^+(\chi))), && \gamma_1^+(\chi) = \rho_+e^{i\xi}+\varepsilon e^{i\chi} \\
    \kappa_2^+(\chi) & = (S_2(\gamma_2^+(\chi)), S_1(\gamma_2^+(\chi))), && \gamma_2^+(\chi) = -\rho_+e^{-\xi}+\varepsilon e^{i\chi}. 
\end{align}

We now proceed with a geometric argument. First, consider the intersection of $(S_1(\gamma_1^+(\xi)), S_2(\gamma_1^+(\xi))$ with the plane spanned by $(s_1, s_2, s_3)$ and $(0,0,1)$ as indicated by the red line in Supplementary Fig. \ref{fig: Contour Example}b. For a point on $(S_1(\gamma_1^+(\chi)), S_2(\gamma_1^+(\chi))$ to land on this plane, we necessarily require a rotation about the axis $(s_2, -s_1, 0)$, and this corresponds to having $\arg(\gamma_1^+(\chi)) = \xi$ or $\xi + \pi$, or correspondingly, $\chi = \xi$ and $\chi = \xi + \pi$. 

As we are concerned with the limit $\varepsilon \searrow 0$, there is no loss of generality in assuming $\varepsilon$ is smaller than $\rho_+$. From this, we have the following. When $\chi = \xi$, $\gamma_1^+(\xi) = (\rho_++\varepsilon)e^{i\xi}$, so the retardance at that point satisfies $\sigma(\rho_++\varepsilon) > \sigma(\rho_+)$. Therefore, $(S_1(\gamma_1^+(\xi)), S_2(\gamma_1^+(\xi)))$ ends up in the direction of $-(s_1, s_2)$ from the origin. This is illustrated by the point $A$ in Supplementary Fig. \ref{fig: Contour Example}b. Conversely, when $\chi = \xi + \pi$, $\gamma_1^+(\xi+\pi) = (\rho_+-\varepsilon)e^{i\xi}$, so the retardance at that point satisfies $\sigma(\rho_+-\varepsilon) < \sigma(\rho_+)$, and $(S_1(\gamma_1^+(\xi+\pi)), S_2(\gamma_1^+(\xi+\pi)))$ ends up in the direction $(s_1, s_2)$ from the origin. This is illustrated in Supplementary Fig. \ref{fig: Contour Example}b by the point $B$. 

From this, we immediately see that the winding number of $\kappa_1^+$ can only be $\pm 1$. To determine the sign, consider a third point $\chi = \xi + \phi$ where $0<\phi<\pi$ and $\lvert \gamma_1^+(\xi + \phi) \rvert = \rho_+$. Notice that $\arg(\gamma_1^+(\xi + \phi)) > \arg(\gamma_1^+(\xi))$. Therefore, the angle made by the axis of rotation in the $s_1-s_2$ plane is more positive when $\chi = \xi + \phi$ as compared to $\chi = \xi$. Therefore, $(S_1(\gamma_1^+(\xi+\phi)), S_2(\gamma_1^+(\xi+\phi)))$ lies in the half-plane given by the dividing line $(s_1t, s_2t)$ not containing $(s_2, -s_1)$. This is illustrated by the point $C$ in Supplementary Fig. \ref{fig: Contour Example}b. From these three distinguished points, it is immediately obvious that the curve $(S_1(\gamma_1^+(\chi)), S_2(\gamma_1^+(\chi)))$ has a winding number of $1$, and so $\kappa_1^+$ has a winding number of $-1$. A similar argument shows that $\kappa_2^+$ also has a winding number of $-1$. 

The $S_3 = -1$ case follows in an analogous manner, where we can likewise show that the winding number of each respective curve is $-1$. As a result, the total contribution of the $S_3 = 1$ and $S_3 = -1$ singularities cancel each other out, and overall, we have a Skyrmion number of $2s_3 = (1+\cos\sigma_{\max})s_3$. 

\subsection{\texorpdfstring{Case III: $\bm{-1 < s_3 < 1,\, 0< \sigma_{\max} < \pi}$}{TEXT}}

Notice first that whenever $0 < \sigma_{\max} < \pi$, the polarization state $s$ is observed only at $r=0$ while the polarization state $-s$ cannot be observed in the output as the angle of rotation that the gradient index lens induces on the Poincar\'{e} is everywhere $<\pi$. This suggests that a change of variables may simplify the problem by reducing the number of singularities that need to be checked. To that end, we define $P = R_c(s)\iota \circ \mathcal{S}$ where $R_c$ is the rotation matrix defined by 
\begingroup \renewcommand{\arraystretch}{1.5}
\begin{equation}
\label{eq: canonical rotation}
    R_c(s) = \begin{pmatrix}
        1-\frac{s_1^2}{1+s_3} & -\frac{s_1s_2}{1+s_3} & s_1 \\ \frac{-s_1s_2}{1+s_3} & 1-\frac{s_2^2}{1+s_3} & s_2 \\ -s_1 & -s_2 & s_3
    \end{pmatrix} \quad \renewcommand{\arraystretch}{1} \text{so that} \quad R_c(s)^Ts = \begin{pmatrix}
        0 \\ 0 \\ 1
    \end{pmatrix}.
\end{equation} \endgroup

If we now consider the potential 
\begin{equation}
    \mathcal{A}_P = P_3 d\left(\tan^{-1}\left(\frac{P_1}{P_2}\right)\right),
\end{equation}
then the singular points of $\mathcal{A}_P$ correspond to $P_3 = (0,0,\pm 1)$ or equivalently $S = \pm s$, and by our remark above, $\mathcal{A}_P$ has just one singularity at $r=0$. Now, to evaluate the integral about $\partial U$, note first that we may write
\begin{align}
    P_3 \frac{\partial}{\partial \theta} \tan^{-1}\left(\frac{P_1}{P_2}\right) & = \frac{P_3}{P_1^2+P_2^2} \left(P_2\frac{\partial P_1}{\partial \theta} - P_1 \frac{\partial P_2}{\partial \theta}\right) = \frac{P_3}{1-P_3^2}\left(P_2\frac{\partial P_1}{\partial \theta} - P_1 \frac{\partial P_2}{\partial \theta}\right) \nonumber  \\
    & = \left(\frac{1}{1-P_3^2}-\frac{1}{1+P_3}\right)\left(P_2\frac{\partial P_1}{\partial \theta} - P_1 \frac{\partial P_2}{\partial \theta}\right)
\end{align}
so that 
\begin{align}
    \frac{1}{4\pi}\int_0^{2\pi} P_3(r_0e^{i \theta}) & \frac{d}{d\theta} \tan^{-1} \left( \frac{P_1(r_0e^{i\theta})}{P_2(r_0e^{i\theta})} \right)  d\theta \nonumber  \\ & = \frac{1}{2}\times \left(\text{winding number of $(P_2(r_0e^{i\theta}), P_1(r_0e^{i\theta}))$ about $(0,0)$}\right) \nonumber  \\
    & \quad - \frac{1}{4\pi}\int_0^{2\pi} \left(\frac{1}{1+P_3(r_0e^{i\theta})}\right)\left(P_2(r_0e^{i\theta})\frac{d P_1}{d \theta}(r_0e^{i\theta}) - P_1(r_0e^{i\theta}) \frac{d P_2}{d \theta}(r_0e^{i\theta})\right) d\theta.
\end{align}

By direct computation, one may show that 
\begin{align}
    \left(\frac{1}{1+P_3(r_0e^{i\theta})}\right) & \left(P_2(r_0e^{i\theta})\frac{d P_1}{d \theta}(r_0e^{i\theta}) - P_1(r_0e^{i\theta}) \frac{d P_2}{d \theta}(r_0e^{i\theta})\right) = -2(1-\cos \sigma_{\max})s_3 \nonumber  \\ & - 2\sin\sigma_{\max}(1-\cos\sigma_{\max}) \frac{(s_1\sin2\theta-s_2\cos2\theta)^3+(1-s_1^2-s_2^2)(s_1\sin2\theta-s_2\cos2\theta)}{1+ \cos\sigma_{\max} + (1-\cos\sigma_{\max})(s_1\cos2\theta+s_2\sin2\theta)^2}
\end{align}
Note further that 
\begin{align}
    s_1\cos2\theta+s_2\sin2\theta & = \sqrt{s_1^2+s_2^2}\cos(2\theta - \phi) \\
    s_1\sin2\theta - s_2\cos2\theta & = \sqrt{s_1^2+s_2^2}\sin(2\theta-\phi)
\end{align}
where $\tan \phi = s_2/s_1$. Now, setting $R_s = \sqrt{s_1^2+s_2^2}$, we have, 
\begingroup
\allowdisplaybreaks
\begin{align}
    - \frac{1}{4\pi}& \int_0^{2\pi} \left(\frac{1}{1+P_3(r_0e^{i\theta})}\right)\left(P_2(r_0e^{i\theta})\frac{d P_1}{d \theta}(r_0e^{i\theta}) - P_1(r_0e^{i\theta}) \frac{d P_2}{d \theta}(r_0e^{i\theta})\right) d\theta \nonumber  \\
    & = (1-\cos\sigma_{\max})s_3+\frac{\sin\sigma_{\max}(1-\cos\sigma_{\max})}{2\pi}\int_0^{2\pi} \frac{R_s^3\sin^3(2\theta-\phi)+(1-s_1^2-s_2^2)R_s\sin(2\theta-\phi)}{1+\cos\sigma_{\max} + (1-\cos\sigma_{\max})R_s^2\cos^2(2\theta-\phi)} d\theta \nonumber  \\
    & = (1-\cos\sigma_{\max})s_3+\frac{\sin\sigma_{\max}(1-\cos\sigma_{\max})}{2\pi}\int_{-\pi}^{\pi} \frac{R_s^3\sin^3(2\theta)+(1-s_1^2-s_2^2)R_s\sin(2\theta)}{1+\cos\sigma_{\max} + (1-\cos\sigma_{\max})R_s^2\cos^2(2\theta)} d\theta \nonumber  \\
    & = (1-\cos\sigma_{\max})s_3,
\end{align}
\endgroup
where the integral in the second last line vanishes by the oddness of the integrand.

Now to account for the singularity at $r=0$, note that $(P_2, P_1)$ is never zero when $r>0$, and therefore, for each $0<r\leq r_0$, $H\colon S^1\times [r, r_0] \longrightarrow \mathbb{R}^2-\{(0,0)\}$ given by 
\begin{equation}
    H(e^{i\theta}, \rho) = (P_2(\rho e^{i\theta}), P_1(\rho e^{i\theta}))
\end{equation}
is a well-defined homotopy from $(P_2(re^{i\theta}), P_1(re^{i\theta}))$ to $(P_2(r_0e^{i\theta}), P_1(r_0e^{i\theta}))$. Consequently, the winding number of $(P_2(re^{i\theta}), P_1(re^{i\theta}))$ about $(0,0)$ is equal to that of $(P_2(r_0e^{i\theta}), P_1(r_0e^{i\theta}))$ about $(0,0)$ for each $r>0$. From this, we have 
\begin{align}
    \text{Skyrmion Number} & = (1-\cos\sigma_{\max})s_3 \nonumber  \\ 
    & \quad \quad - \frac{1}{2}\times \lim_{\varepsilon \searrow 0} \left(\text{winding number of $(P_2(\varepsilon e^{i\theta}), P_1(\varepsilon e^{i\theta}))$ about $(0,0)$} \right) \nonumber  \\
    & \quad \quad + \frac{1}{2}\times \left(\text{winding number of $(P_2(r_0e^{i\theta}), P_1(r_0e^{i\theta}))$ about $(0,0)$}\right) \nonumber  \\
    & = (1-\cos\sigma_{\max})s_3
\end{align}

From this, we see that the formula 
\begin{equation}
    \text{SkyN}(\mathcal{S}) = (1-\cos\sigma_{\max})s_3
\end{equation}
holds in all cases. 

\section{Experiment Assembly}

\begin{figure}[!h]
    \centering
    \includegraphics[width=0.7\textwidth]{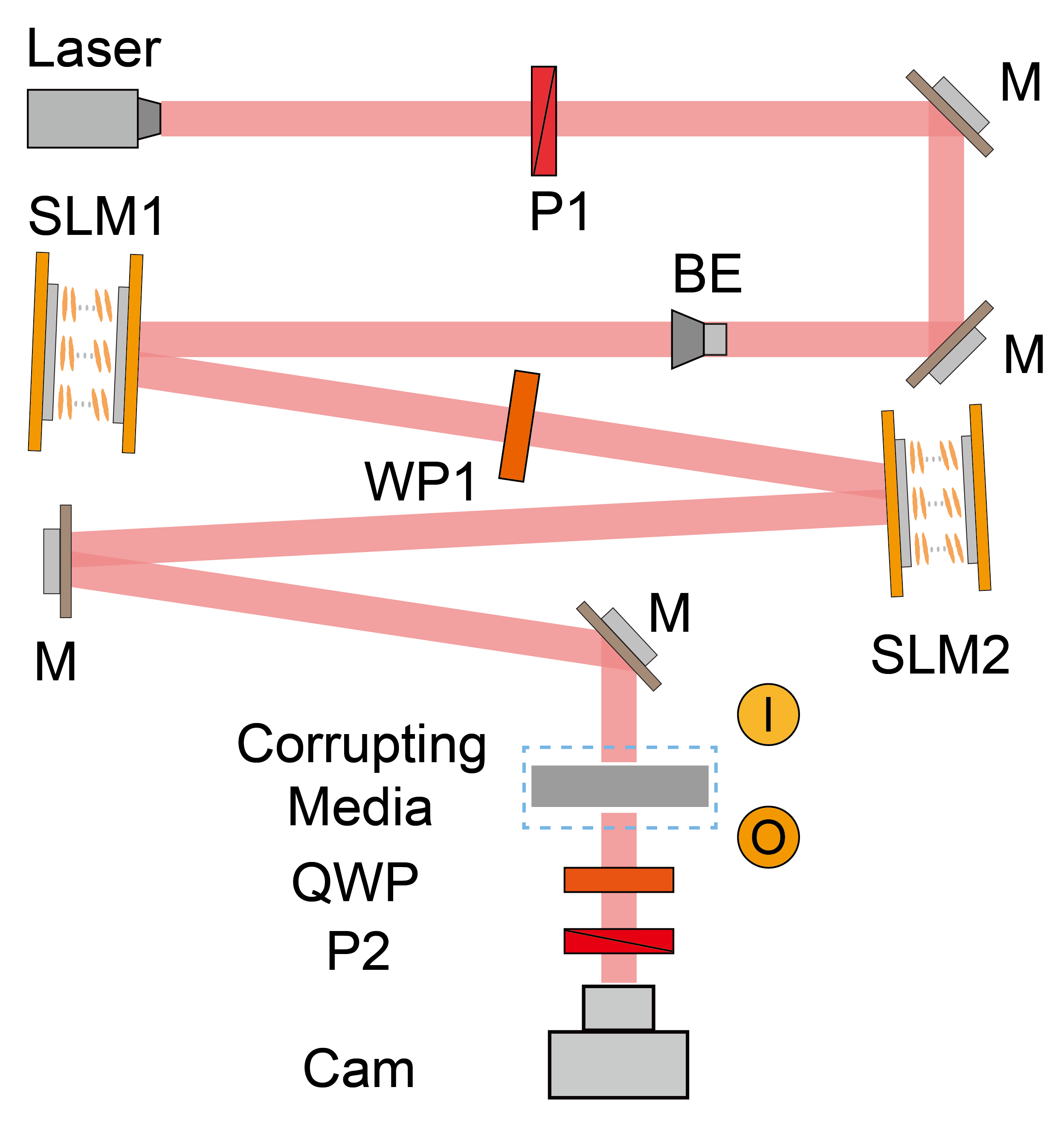}
    \caption{{\bf Experiment assembly.} The complex beam generator and polarimetry configuration adopted in experiments. Components include a He-Ne laser (Melles Griot, 05-LHP-171, 632.8 nm); P1, P2: fixed polarizers (Thorlabs, GL10-A); SLM1, SLM2: spatial light modulators (Hamamatsu, X10468-01); BE: beam expander; WP: fixed waveplate cascade (350-850nm) that achieves a 45\textsuperscript{$\circ$} relative rotation between SLMs; QWP: rotating quarter waveplate (Thorlabs, WPH10M-633) used in polarimetry; Cam: camera (Thorlabs, DCC3240N).}
    \label{fig:setup}
\end{figure}

\newpage

\section{Mueller Matrices of Select Experiment Configurations}

\begin{figure}[!h]
    \centering
    \includegraphics[width=0.95\textwidth]{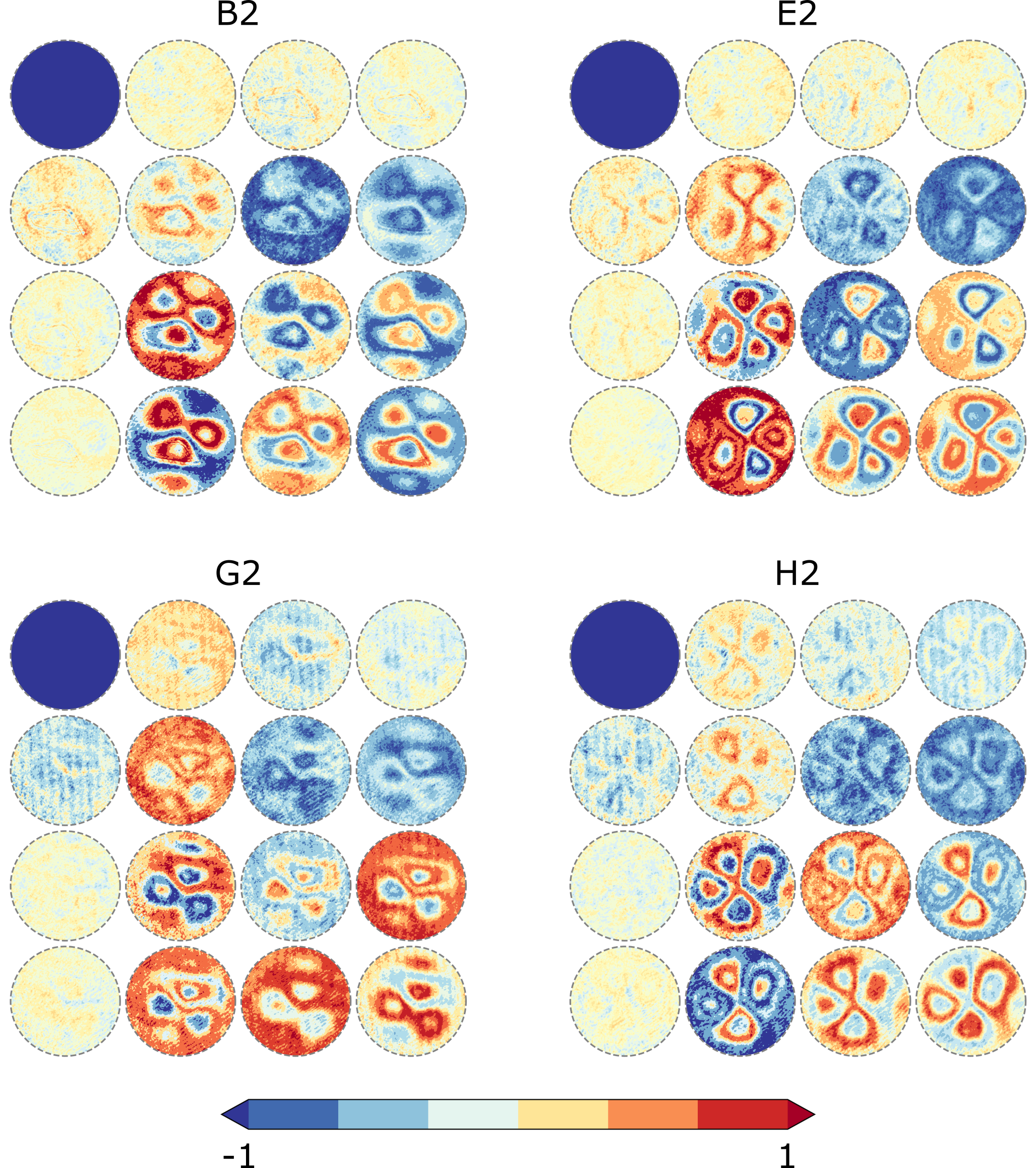}
    \caption{{\bf Mueller matrices of select experiment configurations.} The Mueller matrices of experiment configurations B2, E2, G2, and H2. This figure gives a quantitative characterization of the aberrations and highlights their spatially varying nature.}
    \label{fig:MM}
\end{figure}